\documentclass[fleqn,usenatbib]{mnras}

\usepackage[T1]{fontenc}
\usepackage{ae,aecompl}

\usepackage{eso-pic}

\usepackage{graphicx}	
\usepackage{amsmath}	
\usepackage{amssymb}	
\usepackage[shortlabels]{enumitem}

\usepackage{multicol}
\usepackage[normalem]{ulem}

\title[MaNGA: IMF gradients and $R_*-M_*$]{The half mass radius of MaNGA galaxies: Effect of IMF gradients}

\author[Bernardi et al.]{
  \parbox{\textwidth}{
    M.~Bernardi$^{1}$\thanks{E-mail: \texttt{\rm \texttt{bernardm@sas.upenn.edu}}}, R.~K.~Sheth$^1$, H.~Dom\'{i}nguez S{\'a}nchez$^{2}$, B. Margalef-Bentabol$^{1}$, D. Bizyaev$^{3,4}$ and R. R. Lane$^{5}$}  \\
 \vspace{0.cm}\\~\\
$^{1}$ Department of Physics and Astronomy, University of Pennsylvania, Philadelphia, PA 19104, USA\\
$^{2}$ Institute of Space Sciences (ICE, CSIC), Campus UAB, Carrer de Magrans, E-08193 Barcelona, Spain\\
$^{3}$ Apache Point Observatory and New Mexico State University, P.O. Box 59, Sunspot, NM, 88349-0059, USA\\
$^{4}$ Sternberg Astronomical Institute, Moscow State University, Moscow, Russia\\
$^{5}$ Centro de Investigaci{\'o}n en Astronom{\'i}a, Universidad Bernardo O'Higgins, Avenida Viel 1497, Santiago, Chile\\  
}

\pubyear{2022}

\begin{document}
\label{firstpage}
\pagerange{\pageref{firstpage}--\pageref{lastpage}}
\maketitle

\begin{abstract}
  Gradients in the stellar populations (SP) of galaxies -- e.g., in age, metallicity, stellar Initial Mass Function (IMF) -- can result in gradients in the stellar mass to light ratio, $M_*/L$.  Such gradients imply that the distribution of the stellar mass and light are different.  For old SPs, e.g., in early-type galaxies at $z\sim 0$, the $M_*/L$ gradients are weak if driven by variations in age and metallicity, but significantly larger if driven by the IMF.  A gradient which has larger $M_*/L$ in the center increases the estimated total stellar mass ($M_*$) and reduces the scale which contains half this mass ($R_{e,*}$), compared to when the gradient is ignored. For the IMF gradients inferred from fitting MILES simple SP models to the H$_\beta$, $\langle$Fe$\rangle$, [MgFe] and TiO$_{\rm 2SDSS}$ absorption lines measured in spatially resolved spectra of early-type galaxies in the MaNGA survey, the fractional change in $R_{e,*}$ can be significantly larger than that in $M_*$, especially when the light is more centrally concentrated. The $R_{e,*}-M_*$ correlation which results is offset by 0.3~dex to smaller sizes compared to when these gradients are ignored.  Comparisons with `quiescent' galaxies at higher-$z$ must account for evolution in SP gradients (especially age and IMF) and the light profile before drawing conclusions about how $R_{e,*}$ and $M_*$ evolve. The implied merging between higher-$z$ and the present is less contrived if $R_{e,*}/R_e$ at $z\sim 0$ is closer to our IMF-driven gradient calibration than to unity.

\end{abstract}




\section{Introduction}\label{sec:intro}

Over the last few decades there has been significant interest in the assembly history of galaxies that today are red and dead.  In particular, the size-luminosity correlation of these galaxies is extremely tight \citep{Bernardi2003} as is the size-stellar mass correlation \citep{Shen2003}.  This tightness makes this relation well-suited for evolution studies.  
Evolution in the size-mass correlation is expected.
E.g., some models postulate minor dry mergers at late times which increase sizes without significantly changing the stellar mass \cite[e.g.][]{Hilz2013, Hirschmann2015}.  In others still, star formation quenches ``inside-out'' \cite[the central bulge quenches before the outskirts, e.g.,][]{Nelson2016}, resulting in color gradients; mass- or size-dependent quenching are additional complications.

Quiescent galaxies at high redshift appear to be smaller than quiescent galaxies at $z\sim 0$ of the same or comparable stellar mass:  when galaxy size is plotted versus stellar mass, then the high $z$ population is offset towards smaller sizes \citep{Daddi2005, Buitrago2008, Cimatti2008, vanDokkum2008, vanderWel2014, Chan2016, Barro2017, Mowla2019}.

However, for this to properly constrain models, one must first address the question of how the stellar masses and sizes are estimated. While there has been significant discussion of the difficulty of doing this in the high-$z$ population and of possible systematic effects causing the rapid evolution in $R_e$ (e.g. redshift-dependent selection effects, systematic uncertainties and/or progenitor bias,  \citealt[e.g.][]{vanDokkumFranx1996, vanderWel2009, Shankar2015, Zanisi2021}), there is a potential systematic in the low-$z$ population that has not received much attention, which we study here.

The systematic derives from the fact that galaxies show stellar population gradients.  These gradients affect how one converts from the projected surface brightness profile one observes to the stellar mass profile which one uses to estimate a galaxy's total stellar mass and size.  In particular, if $I(R)$ and $\Upsilon(R)$ denote the surface brightness and stellar mass-to-light ratio at (not within) projected radius $R$, then the total luminosity and stellar mass are
\begin{equation}
 L \equiv 2\pi \int dR\,R\,I(R)\quad{\rm and}\quad 
 M_* \equiv 2\pi \int dR\,R\,I_*(R),
 \label{eq:totM*}
\end{equation}
where $I_*(R) \equiv I(R)\,\Upsilon(R)$.  One may, of course, define a global $\Upsilon_*\equiv M_*/L$ using the total mass and total light, however, what to use for the sizes is more subtle.  Typically, the `effective radius' $R_e$ that is quoted is that $R$ which contains half the projected light:
\begin{equation}
 \frac{L}{2} \equiv 2\pi \int_0^{R_e} dR\,R\,I(R).
\end{equation}
But if $\Upsilon(R)$ depends on $R$, then this $R_e$ will not be the same scale as $R_{e,*}$, the scale which contains half the projected stellar mass:
\begin{equation}
 \frac{M_*}{2} \equiv 2\pi \int_0^{R_{e,*}} dR\,R\,I(R)\,\Upsilon(R).
 \label{eq:R*}
\end{equation}
Note that the difference between $R_e$ and $R_{e,*}$ depends {\em both} on $\Upsilon(R)$ {\em and} on the shape of $I(R)$.  

It has been known for some time that the stellar population in most galaxies varies with $R$.  This will produce $\Upsilon(R)\ne\Upsilon_*$, and hence $R_e\ne R_{e,*}$.  It is useful to think of such $\Upsilon$ gradients in a galaxy as arising from two effects:\\
 (i) variations in age and chemical composition which would arise even if the IMF were the same throughout the galaxy; and\\
(ii) the additional effect of IMF gradients.

There has been previous work studying the difference between $R_e$ and $R_{e,*}$ when the IMF is assumed to be constant.  Most of these are motivated by the fact that the half-light radius has long been known to depend on wavelength \citep{Bernardi2003, LaBarbera2009, Kennedy2015} -- so $R_e$ in most bands is guaranteed to be different from $R_{e,*}$.  However, wavelength dependence in $R_e$ implies color gradients, and colors are a crude proxy for $\Upsilon$.  E.g., \cite{Szomoru2013} use $u-g$ color profiles, and an assumption about how color traces $\Upsilon$, to conclude that $R_{e,*}$ is about 25\% smaller than $R_e$ in massive galaxies over $0.5<z<2.5$.  A similar (optical color) based analysis of a much larger sample ($\sim 7000$ objects compared to $\sim 200$) over $1<z<2.5$ \citep{Suess2019} finds a larger difference.  However, optical colors are {\em not} sensitive to the IMF, so they cannot address effect (ii).  More recently, \cite{Ibarra-Medel2022} have estimated stellar population gradients from spatially resolved spectra (rather than colors) in MaNGA galaxies at $z\sim 0$, and combined them with the archaeology approach of \cite{Lacerna2020} to predict how $R_{e,*}/R_e$ might have evolved.  This approach also does not account for the possibility that IMF gradients may contribute to $\Upsilon$ gradients, and hence to the difference between $R_{e,*}$ and $R_e$. 

This matters because there is growing evidence that the IMF in early-type galaxies is not universal \cite[][and references therein]{Smith2020}.  The IMF in the central regions of early-type galaxies at $z\sim 0$ differs from that in the outskirts (e.g. \citealt{MartinNavarro2015, vD2017, Parikh2018, LaBarbera2018, LaBarbera2019}; but see, e.g., \citealt{Vaughan18, FeldmeierKrause2021}).  If the stellar Initial Mass Function (IMF) is constant within a galaxy, then changes in age, metallicity, element abundances and so on will produce changes in $\Upsilon$, but, for early-type galaxies at $z\sim 0$, these are known to be small \cite[e.g.][]{Mehlert2003, Spolaor2009, Tortora2011, Kuntschner2015, Li2018, califa2019, Ge2021}.  However, changes in the IMF across the galaxy can produce more significant changes in $\Upsilon$.  \citet{Bernardi2018b} argued that IMF-driven gradients in $\Upsilon$ can have profound consequences for how one estimates galaxy stellar masses from stellar populations ($M_*^{\rm SP}$) or from dynamical methods ($M_*^{\rm dyn}$).  In particular, they noted that IMF-driven gradients can bring $M_*^{\rm dyn}$ and $M_*^{\rm SP}$ into agreement, not by shifting $M_*^{\rm SP}$ upwards as advocated by some studies \cite[e.g.][]{Cappellari2013a, Li17}, but by revising $M_*^{\rm dyn}$ estimates in the literature downwards \cite[this is true whether or not the mass contributed by the gradient is a distinct dynamical component, see Appendix~E of][]{Marsden2022}. Recent analyses of quiescent galaxies in the MaNGA survey have shown that IMF gradients do appear to be driving non-negligible $R$ dependence in $\Upsilon(R)$ \citep{DS2019, Bernardi2019, DS2020}. 

In addition to $\Upsilon$ variations within galaxies, it has also been known for some time that $I(R)$ varies systematically across the quiescent galaxy population.  This variation is often quantified by fitting $I(R)$ to a S{\'e}rsic profile \citep{Sersic1963}.  This profile has three free parameters: an amplitude $I_e$, a scale $R_e$ and a shape parameter $n$.  So, for the same $\Upsilon(R)$ profile, it is reasonable to expect the ratio $R_e/R_{e,*}$ to depend on $n$.  Since, in practice, $\Upsilon(R)$ also varies across the population, it is not obvious how different the $R_e-M_*$ and $R_{e,*}-M_*$ relations will be when one accounts for stellar population gradients. The main goal of the present work is to quantify this difference in the early-type galaxy population at $z\sim 0$.\footnote{One might reasonably expect that systems with a ‘bulge’ and a ‘disk’ have different stellar populations, and hence $M_*/L$ gradients, even when the IMF is fixed (e.g. to Kroupa) for both components. Therefore, we study spirals in a separate paper.}

Galaxy structure (encoded in $n$) and stellar populations (age, metallicity, IMF, etc.) also correlate with galaxy morphology.  Since morphology may correlate with assembly history, in what follows, we study the size-stellar mass correlations separately for elliptical slow rotators, elliptical fast rotators and S0s. This means that, to address the effect of IMF-driven gradients on the $R_{e,*}-M_*$ relation, one requires reliable morphology, photometry, and stellar population gradient information for a large sample.  Section~\ref{sec:data} describes the dataset we use, and the associated morphology, size and stellar population estimates.  Section~\ref{sec:results} shows our results.  Section~\ref{sec:hiz} compares our low-$z$ analysis with estimates at higher $z$, and a final section summarizes.  Although we concentrate on the size-mass correlation, an understanding of how $R_{e,*}/R_e$ evolves also impacts studies of the evolution of the stellar mass Fundamental Plane \citep{Bernardi2020, deGraaff2021}.  We leave this to future work.

Estimating the IMF is difficult, with the potential for systematic effects to compromise both the measurement and the modeling/interpretation steps.  A number of these systematics, and the reasons for our particular fiducial choices, are discussed in a companion paper \citep{Bernardi2022}. There we show that, although our fiducial choice of stellar population model, MILES+Padova \citep{Vazdekis2015, Pietrinferni2013, DS2019}, results in large IMF-driven $M_*/L$ gradients, other choices sometimes do and sometimes do not result in similar gradients.  Thus, the results which follow are only as good as the fiducial stellar population models we use to estimate $M_*/L$.

\section{Data}\label{sec:data}
Our study, which requires reliable morphology, photometry, and stellar population gradient information for a large sample, is made possible by the MaNGA survey \cite[e.g.][]{Bundy2015, Aguado2019, Westfall2019}. In MaNGA, morphology, photometry, and stellar population gradient information are available on a galaxy-by-galaxy basis, but because the stellar population gradients require higher signal-to-noise spatially resolved spectroscopy than is available for single objects, we estimate these from stacking spectra of objects having similar properties.  The next subsections describe the survey, the morphological and photometric parameters we use, as well as how our stacked spectra were defined.

\subsection{MaNGA survey and photometry}
The MaNGA survey (\citealt{Bundy2015,Drory2015,Law2015,Law2016,Yan2016a, Yan2016b}), which is a component of the Sloan Digital Sky Survey IV (\citealt{Gunn2006, Smee2013, Blanton2017}; hereafter SDSS IV), uses integral field units (IFUs) to measure spectra across $\sim$ 10000 nearby galaxies.

The MaNGA final data release (DR17 -- \citealt{sdssDR17}) includes spectra (wavelength coverage $3500-10^4$\AA) of $\sim 10^4$ nearby ($0.03<z<0.15$) galaxies distributed across 4000~deg$^2$ and uniform over the mass range $M_* \ge 10^9-10^{12}\, M_\odot$ with no size, inclination, morphology or environmental cuts.  Two-thirds of the sample has spatial coverage, at about 1~kpc resolution, to 1.5 times the half-light radius of a galaxy; the other third of the sample has coverage to 2.5$R_e$.  Early-type galaxies make up about thirty percent of the sample. 
The MaNGA selection function, while complicated, is well defined \citep{Wake2017}.  In what follows, when necessary, we use {\tt ESWEIGHT}, which is provided and recommended by \cite{Wake2017} as a way to crudely account for this selection.

In this work, we use morphological classifications and photometric parameters from two Value Added Catalogs which are included in the final MaNGA DR17 data release (the {\tt DR17 MaNGA PyMorph photometric} (DR17-MPP-VAC) and {\tt DR17 MaNGA Morphology Deep Learning} (DR17-MMDL-VAC) catalogs; see \citealt{DS21} for details). These catalogs are updated/completed versions of the corresponding MaNGA DR15 VACs \citep{Fischer2019} and include 10,293 entries which correspond to 10,127 unique galaxies. To reduce aperture and evolution effects, \cite{DS2019} recommend limiting the sample to $z\le 0.08$.  However, in their analysis of the stellar populations of these objects, \cite{Bernardi2022} find that including objects to $z\le 0.15$ makes no difference -- it just improves the statistics. We have checked that this is also true of all the analysis which follows, so the results we show include objects with $z\le 0.15$.

The DR17-MPP-VAC provides photometric parameters from single component S\'ersic (Ser) and two-component S\'ersic + Exponential (SerExp) fits to the 2D surface brightness profiles of the MaNGA DR17 galaxy sample in the SDSS $g$, $r$, and $i$ bands. In addition to total magnitudes, effective radii, S\'ersic indices, axis ratios $b/a$, etc., this VAC also includes a flagging system (FLAG$\_$FIT) which indicates the preferred fit model:
\begin{itemize}
  \item FLAG$\_$FIT$=$1 means that Ser fit is preferred (the SerExp fit may be unreliable);
  \item FLAG$\_$FIT$=$2 means that the SerExp fit is preferred (the Ser fit may be unreliable);
  \item FLAG$\_$FIT$=$0 means that both Ser and SerExp fits are acceptable; and
  \item FLAG$\_$FIT$=$3 means that none of the fits were reliable and so no parameters are provided.
\end{itemize}
For each galaxy, we use the best-fit parameters in the SDSS $r$-band for the model indicated by FLAG$\_$FIT. When  FLAG$\_$FIT = 0 -- i.e.,  no preference between Ser or SerExp fits -- we use the values returned by the latter. (The results which follow do not depend on this choice.) In what follows, we use the `truncated' magnitudes and sizes if not otherwise specified. Also $R_e$ is the half-light radius of the truncated profile along the semimajor axis, i.e. $R_e = R_{e,{\rm maj}}$ (not circularized).

\begin{figure}
  \centering
  \includegraphics[width=0.9\linewidth]{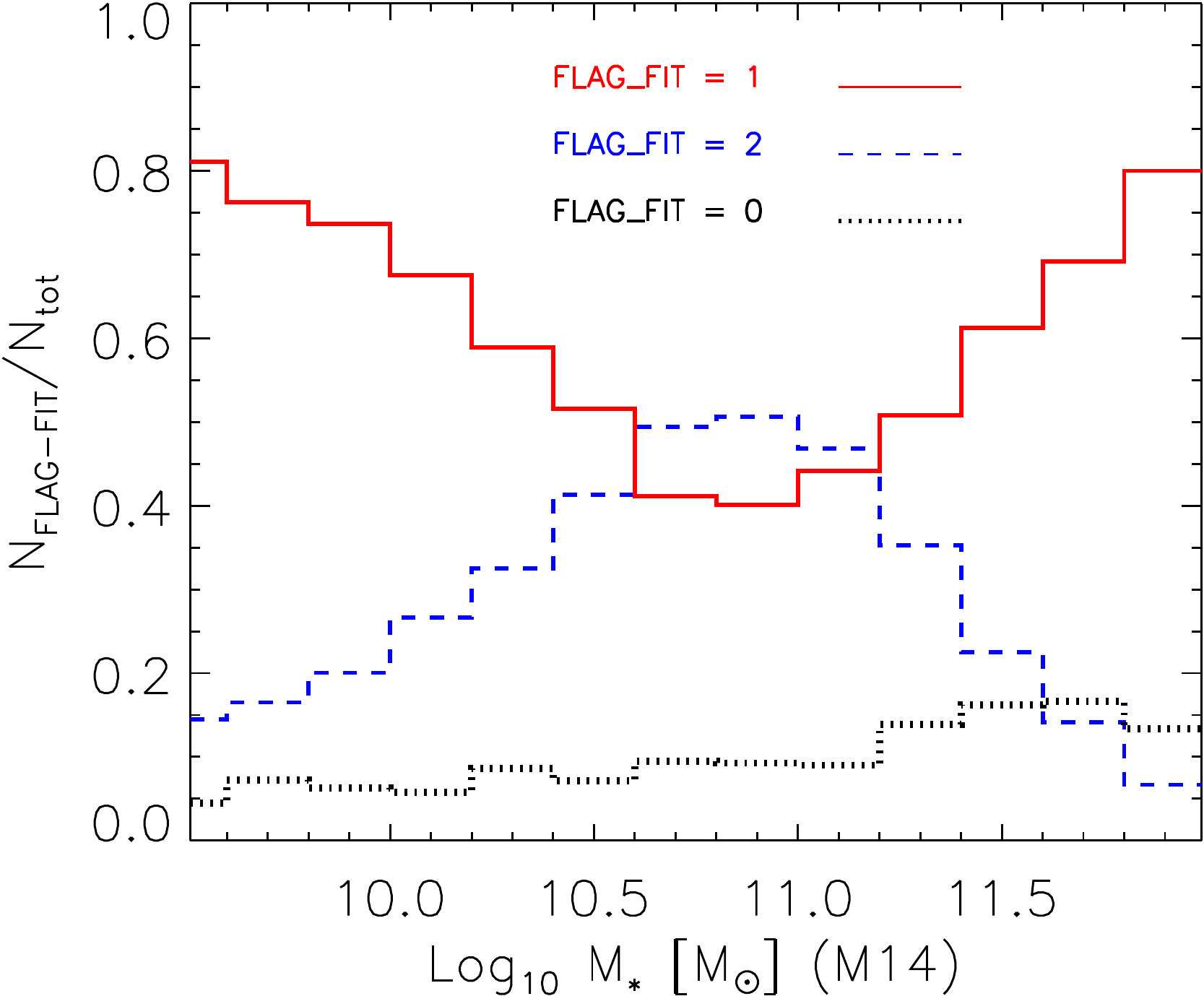}
  \caption{Distribution of stellar masses (from combining $M_*/L$ of M14 with the DR17-MPP-VAC estimate of $L$) for objects which are better fit by a single S{\'e}rsic profile, a two-component SerExp profile or for which both are equally acceptable ({\tt FLAG\_FIT = 1, 2} and {\tt 0}, respectively).  Note that two components are required at intermediate masses.}
  \label{fig:flagM*}
\end{figure}

\begin{figure}
  \centering
  \includegraphics[width=0.9\linewidth]{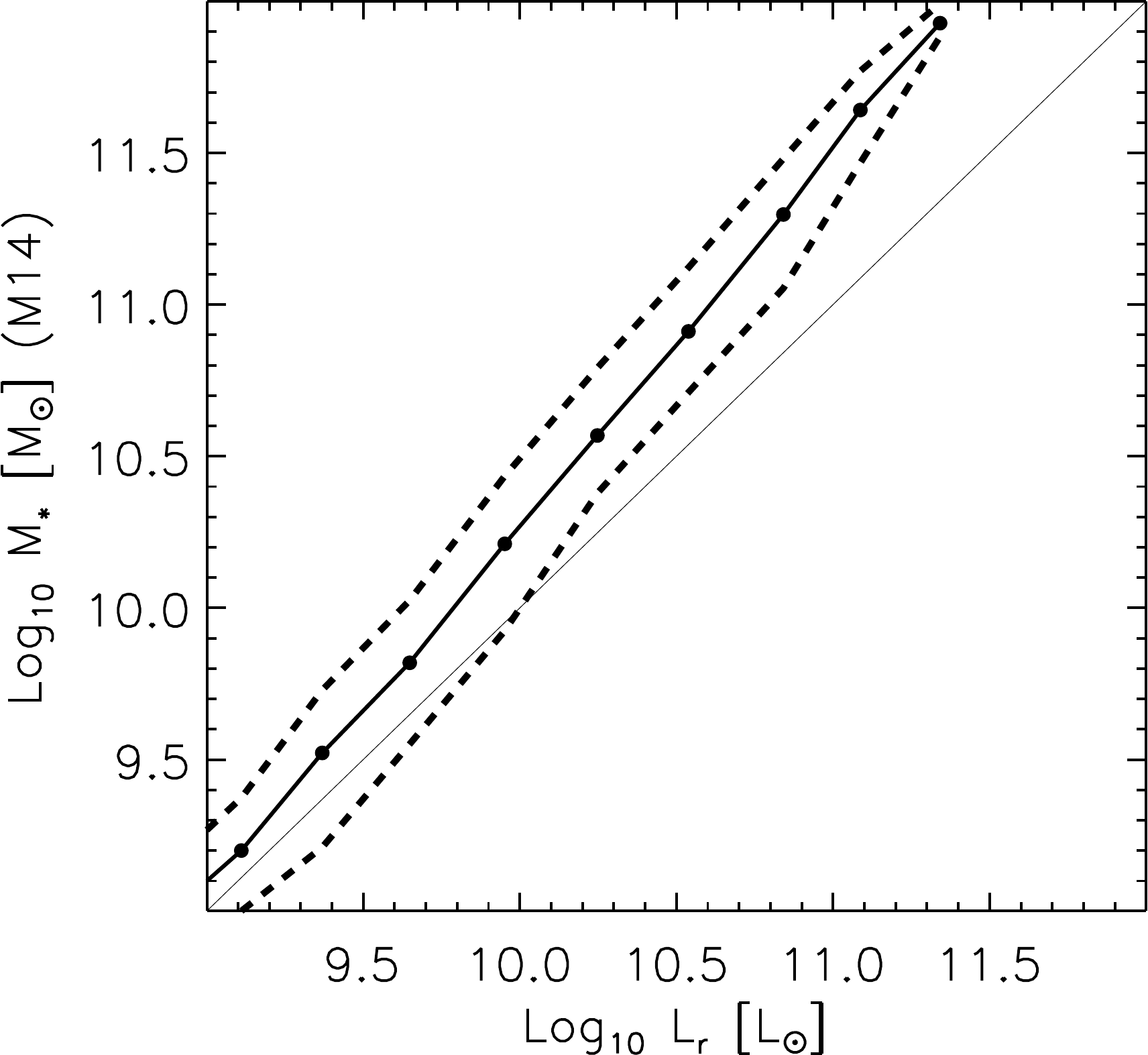}
  \caption{Correlation between $M_*$ and $L_r$ for the objects in our sample which have $M_*/L$ estimates from M14.  Solid line shows the median $M_*$ in bins of $L_r$; dashed lines show the region which encloses 68\% of the objects around the median.}
  \label{fig:M*L}
\end{figure}

To provide some intuition into the FLAG$\_$FIT values, Figure~\ref{fig:flagM*} shows that the least massive and most massive objects tend to have FLAG$\_$FIT$=$1 (i.e. single component S{\'e}rsic fit is preferred), whereas objects of intermediate masses tend to be two-component systems.  For each object, $M_*$ comes from combining the $M_*/L$ estimate of \citet[][hereafter M14]{Mendel2014} with the value of $L$ in the VAC that is appropriate for the FLAG$\_$FIT value.  \cite[We shift the M14 values from a Chabrier to a Kroupa IMF using the values provided in Table~2 of][]{Bernardi10}.  For about 20\% of the sample, M14 $M_*/L$ values are not available.  For these, we estimate $M_*$ from the $M_*-L$ correlation shown in Figure~\ref{fig:M*L}, which is defined by the objects in our sample for which both $M_*/L$ and $L$ are available.  The relation is tight, so this should be a reasonable approximation to the actual $M_*$ value.

\begin{figure*}
  \centering
  \includegraphics[width=0.9\linewidth]{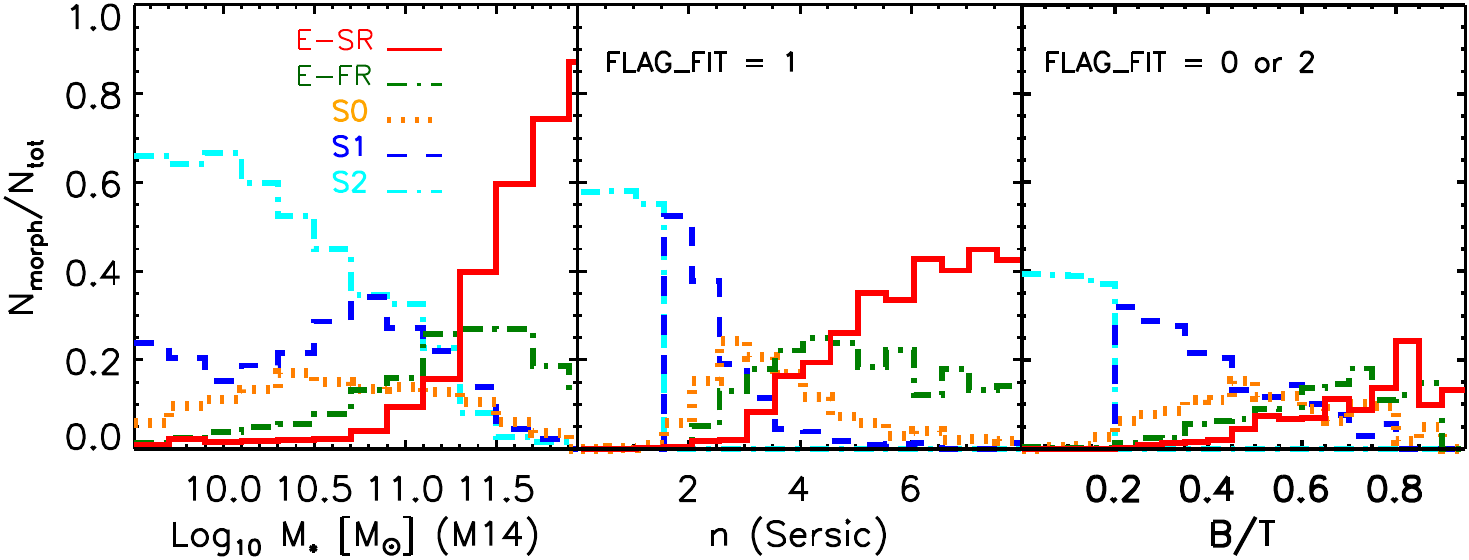}
  \caption{Left: Distribution of stellar mass for objects of different morphology:  slow rotating Ellipticals (E-SR) dominate above $10^{11.5}M_\odot$ and Spirals with B/T$<0.2$ or $n<1.5$ (S2) dominate below $10^{10.5}M_\odot$.
    Middle:  Distribution of S{\'e}rsic index $n$ for objects with {\tt FLAG\_FIT = 1}:  E-SRs dominate at $n>5$ and Spirals dominate at $n<2$; E-FRs have a broad distribution around $n=4$; S0s have a narrow distribution around $n=3$.
    Right:  Distribution of bulge to total ratio B/T for objects with {\tt FLAG\_FIT = 2} or {\tt 0}:  E-SRs dominate at B/T$>0.7$; Spirals dominate at low B/T ($<0.4$); E-FRs have a nearly uniform distribution above B/T$=0.4$; S0s have a broad distribution centered on B/T$\sim 0.5$.}
  \label{fig:morphM*}
\end{figure*}

The DR17-MMDL-VAC provides morphological classifications, such as e.g. the T-Type parameter \cite[][which correlates with Hubble-type]{Ttype1959}, P$_{\rm LTG}$ which separates early-type from late-type galaxies, P$_{\rm S0}$ which separates pure ellipticals (E) from S0s, etc. We refer the reader to \citet{DS21} for further details.
We further subdivide the Es into slow (E-SR) and fast rotators (E-FR) based on the ellipticity and the spin $\lambda_e$ \cite[as in][]{Emsellem2007} (we have corrected $\lambda_e$ for seeing following \citealt{Graham2018} and we refer to the corrected value as $\lambda_{\rm e-PSF}$), and the Spirals into objects which have a small bulge-to-total light ratios (B/T < 0.2) or S{\'e}rsic $n$ (< 1.5), i.e. bulgeless galaxies, vs slightly higher bulge fractions or S{\'e}rsic $n$.

Briefly, for this work, we select all objects with FLAG$\_$FIT $\ne 3$ and, to exclude repeated observations, we choose DUPL$\_$ID $\le 1$. We also exclude galaxies for which a visual inspection of the spectra showed contamination by neighbours. 
The selected objects were classified as follows:
\begin{itemize}
\item E-SR:  T-Type $\le 0$ AND P\_LTG $< 0.5$ AND P\_S0 $\le 0.5$ \\
  AND VISUAL\_CLASS = 1 AND $\epsilon \le 0.4$ AND $\lambda_{\rm e-PSF} \le  0.08+\epsilon/4$.
  This resulted in 730 objects;
 \item E-FR: Similar but $\lambda_e > 0.2$.
   This resulted in 698 objects (note that here we use $\lambda_e$, not $\lambda_{\rm e-PSF}$ since the PSF correction tends to increase the value of $\lambda_e$; so we prefer purity to completeness. Using $\lambda_{\rm e-PSF}$ would have resulted in 973 objects. The excluded galaxies are distributed homogenously along the $R_e - M_r$ and $\sigma_0 - M_r$ relations, so this selection does not introduce selection effects into the stacking analysis which we describe below.)
\item S0: T-Type $\le 0$ AND P\_LTG $< 0.5$ AND P\_S0 $> 0.5$ \\
   AND VISUAL\_CLASS $= 2$.
   This resulted in 751 objects. Distinguishing between S0 and Sa is not easy \cite[for details, see Section 3.4.1 of][]{DS21}.  In this case also, we prefer purity to completeness.
\end{itemize}
Although not the main focus of this study, for future work we separate Spirals into:
\begin{itemize}
\item S1: T-Type $> 0$ AND P\_LTG $\ge 0.5$ AND VISUAL\_CLASS $= 3$ AND \{[(FLAG$\_$FIT = 0 OR FLAG$\_$FIT = 2) AND B/T $> 0.2$] OR [FLAG$\_$FIT = 1 AND $n$ $> 1.5$]\}, i.e. these are spirals with higher bulge fractions or S{\'e}rsic index $n$. This results in 1481 objects.  
\item S2: Similar but \{[(FLAG$\_$FIT = 0 OR FLAG$\_$FIT = 2) AND B/T $\le 0.2$] OR [FLAG$\_$FIT = 1 AND $n$ $\le 1.5$]\}, i.e. these are spirals which have a small B/T or small S{\'e}rsic index. This results in 3107 objects.
 
\end{itemize}

Figure~\ref{fig:morphM*} shows the distribution of $M_*$, $n$ and B/T for the five morphological types.  Evidently, E-SRs tend to have the largest masses, S{\'e}rsic indices and B/T values, and S2 spirals have the smallest. S0s tend to have $n\sim 3\pm 1$ and B/T$\sim 0.5\pm 0.3.$  These are sensible trends.

\subsection{Stellar population parameters from stacked MaNGA spectra}\label{stacks}

Previous analyses of IMF gradients have been based on a handful of objects \cite[e.g.][]{MartinNavarro2015, LaBarbera2017, vD2017, Vaughan18, LaBarbera2019, FeldmeierKrause2021}. The samples are small in part because determining the IMF is not easy: changes in the IMF only lead to rather subtle effects on the spectrum, some of which are degenerate with other stellar population differences (e.g., star formation histories, chemical abundances, etc.). Crudely speaking, this is because the IMF-sensitive features in the spectrum are due to dwarf stars which do not dominate the total light, especially in the optical. High signal-to-noise spectra ($> 100$) are required to disentangle IMF gradients from these other effects. 

Whereas individual spectra in the central region of a MaNGA galaxy have S/N $\sim 100$, the vast majority of the spaxels have S/N $< 50$.  Fortunately, the MaNGA sample is large enough that one can reach S/N $>100$ by stacking spectra of similar objects, even after subdividing each bin in morphology (E-SR, E-FR and S0), into bins in luminosity $L$, central velocity dispersion $\sigma_0$ and radial distance $R$ \citep{DS2019, Bernardi2019, DS2020}.

Therefore, to estimate stellar population parameters, \cite{Bernardi2022} stacked the spectra of the DR17 MaNGA sample in a similar way.  Briefly, for each morphological type, they separated galaxies into luminosity bins, which run from $M_r=-19.5$ to $-23.5$~mags in steps of 1~mag.  In each luminosity bin, they further subdivided the objects based on the velocity dispersion reported in the MaNGA database as having been measured within 0.1 of the half-light radius (see Table~\ref{tab:Cat}). They measured a number of line-index strengths in the stacked spectra.
They then used the MILES+Padova\footnote{Following \cite{DS2019}, these were obtained by starting with the MILES models of \cite{Vazdekis2015} with the Padova isochrones of \cite{Girardi2000}, and correcting for $\alpha$-element abudances using the BaSTI isochrones of \cite{Pietrinferni2006}.} simple stellar population (SSP) models to estimate age, [M/H], [$\alpha$/Fe] and IMF profiles, by fitting the (emission corrected) H$_\beta$, $\langle$Fe$\rangle$, TiO$_{\rm 2SDSS}$ and [MgFe] line strengths measured in the stacked spectra.

In the MILES models, a `bimodal IMF' is defined by a single parameter $\Gamma_b$ that controls both the power law slope at the high-mass end, and the way in which it flattens at lower masses \citep{Vazdekis1996}.  In effect, $\Gamma_b$ controls the dwarf-to-giant ratio in the IMF.  
The Kroupa Universal IMF is closely approximated by a bimodal IMF with $\Gamma_b = 1.35$; more bottom-heavy IMFs have larger $\Gamma_b$.
For our purposes, the $\Gamma_b$ parametrization is sufficiently general, as most IMF-sensitive features depend on the dwarf-to-giant ratio \citep[e.g.,][]{LaBarbera2013, LaBarbera2016}, and not on the detailed shape of the IMF.

When fitting, the models span 
 1–14 Gyr in age, 
 $-0.7$ to $0.2$~dex in [M/H], 
 $0$ to $0.4$ in [$\alpha$/Fe] and 
 1.3–3.5 for the IMF slope parameter $\Gamma_b$. 
Typically, the bimodal IMF parameter $\Gamma_b$ is $\approx 1.35$ (Kroupa) on large scales, but it increases towards the center.
This variation in $\Gamma_b$, along with associated self-consistent variations in age, metallicity and $\alpha$-enhancement, (the $\Gamma_b$ determination is mainly degenerate with age) gives rise to a gradient in $M_*/L$.
In what follows, we will use these $M_*/L$ gradients to illustrate the effects on the size-$M_*$ relation. \cite{Bernardi2022} show that although the statistical errors on the derived SSP parameters and associated $M_*/L$ gradients are negligible, they are model dependent. \cite{Bernardi2022} and \cite{DS2019} discuss why this SSP model and these four absorption lines are expected to be reliable.

\section{Results}\label{sec:results}

\subsection{Expected consequences of gradients}

Before using the actual gradients in MaNGA, we use the following simple model to build intuition.  This model assumes that $\Upsilon \equiv M_*/L$ decreases linearly from its value $\Upsilon_0$ at $R=0$ until some $R=R_{\rm flat}$ after which it is constant and equal to $\Upsilon_\infty$:
\begin{equation}
 \Upsilon(R) = \Upsilon_0  - (\Upsilon_0 - \Upsilon_\infty)\,(R/R_{\rm flat})\quad {\rm when}\quad R<R_{\rm flat}.
 \label{eq:toy}
\end{equation}
The only free parameters are the scale $R_{\rm flat}$, the value $\Upsilon_\infty$ and the ratio $\Upsilon_\infty/\Upsilon_0$.  This expression, when inserted in equations~(\ref{eq:totM*}) and~(\ref{eq:R*}) yields $M_*$ and $R_{e,*}$.  We do this for each MaNGA galaxy, so that we include the correlations between galaxy structure (S{\'e}rsic $n$), size ($R_e$), and $\Upsilon_\infty$.

\begin{figure}
  \centering
  \includegraphics[width=0.9\linewidth]{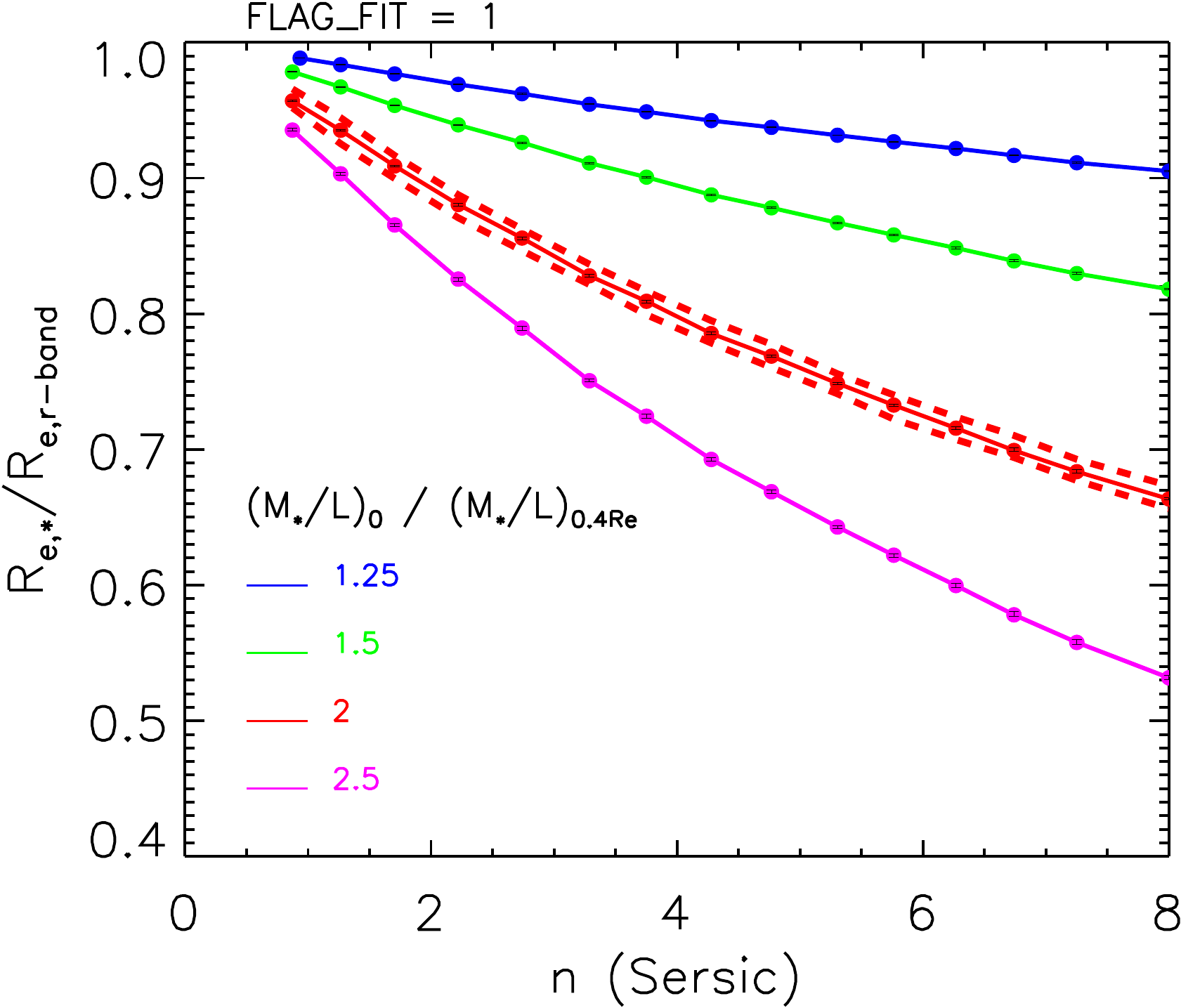}
  \caption{Expected dependence of the ratio of half-mass to half-light projected radius on S{\'e}rsic $n$, for objects with {\tt FLAG\_FIT = 1}, as the assumed $M_*/L$ gradient ($\Upsilon_0/\Upsilon_\infty = (M_*/L)_0/(M_*/L)_{0.4R_e}$) is varied from weak (top) to strong (bottom).
    Dashed curves show the region which encloses 68\% of the objects around the median $R_{e,*}/R_e$. Statisical errors, shown as black bars, are smaller than the symbol sizes. The same $M_*/L$ gradient has a much larger effect if $n$ is large. }
  \label{fig:ratio-n}
\end{figure}

\begin{figure}
  \centering
  \includegraphics[width=0.9\linewidth]{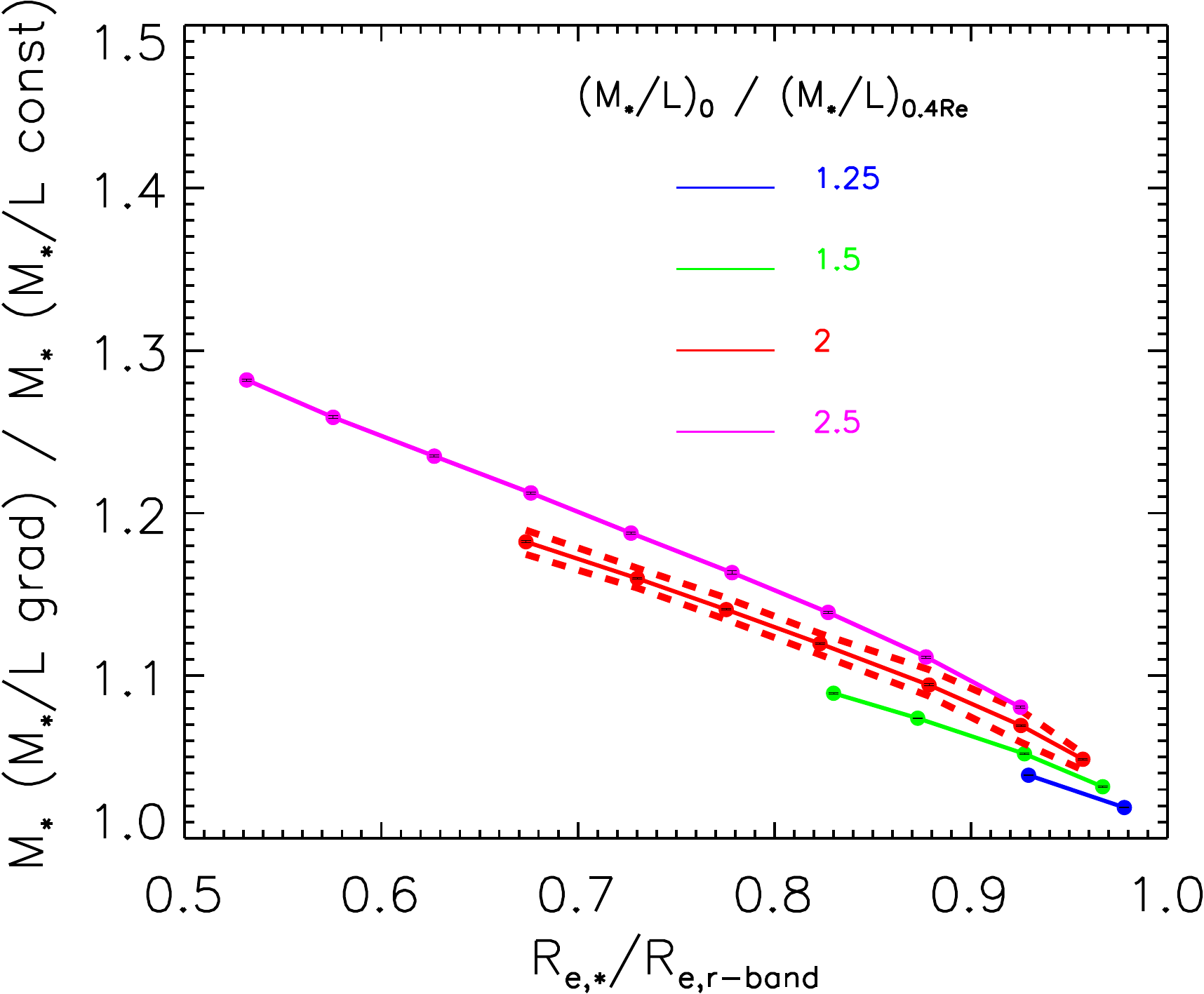}
  \caption{Fractional change in mass versus corresponding change in radius for a few choices of gradient strength.  Gradients increase the mass but decrease the half-mass radius.}
  \label{fig:alphaSize}
\end{figure}

\begin{figure}
  \centering
  \includegraphics[width=0.9\linewidth]{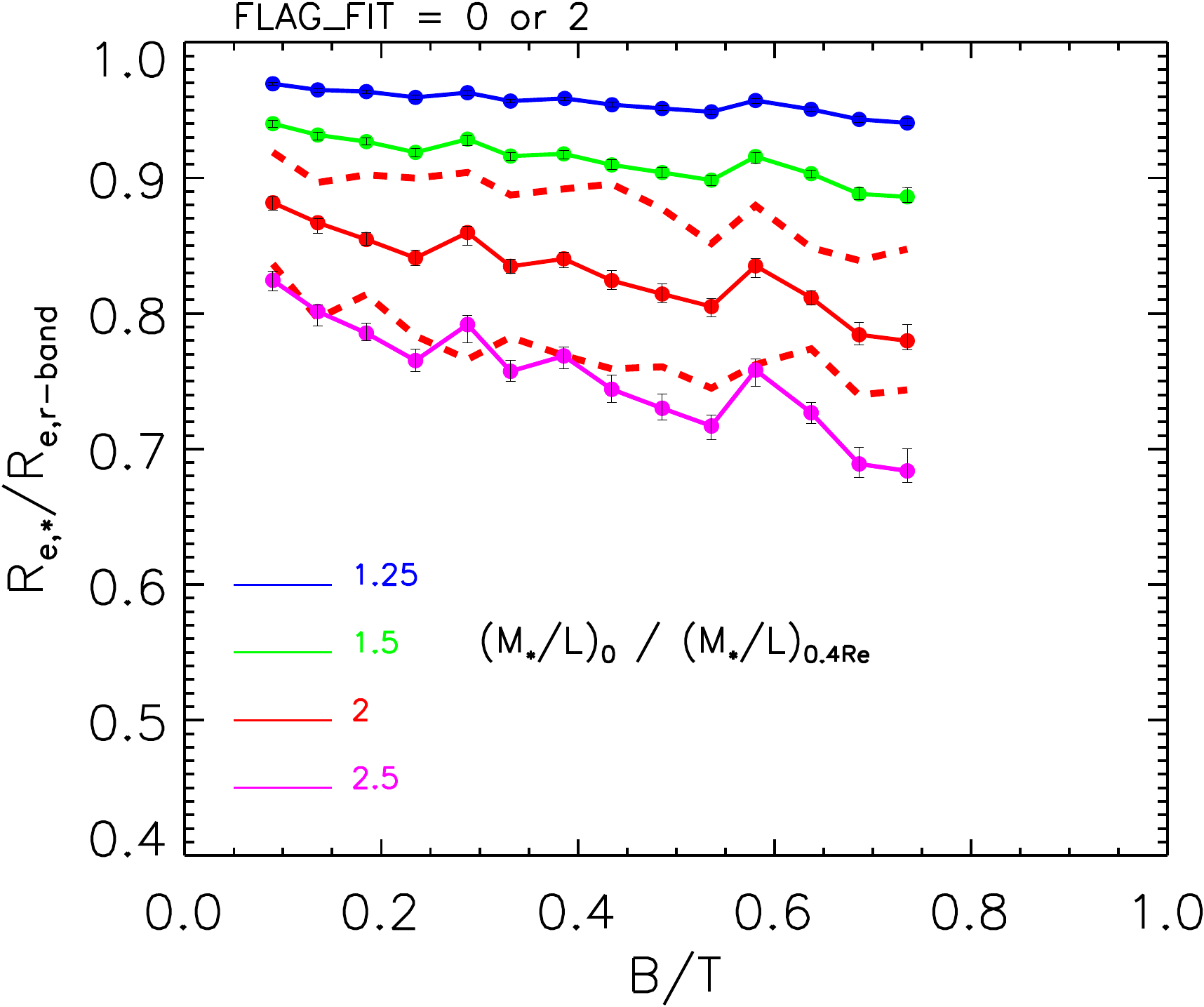}
  \caption{Same as Figure~\ref{fig:ratio-n}, but $R_{e,*}/R_e$ is shown as a function of B/T for the subset of objects with {\tt FLAG\_FIT = 2} or {\tt 0}.  The same $M_*/L$ gradient has a slightly larger effect if B/T is large, but the dependence on B/T is not as large as that on $n$.}
  \label{fig:ratio-BT}
\end{figure}

\begin{figure}
  \centering
  \includegraphics[width=0.9\linewidth]{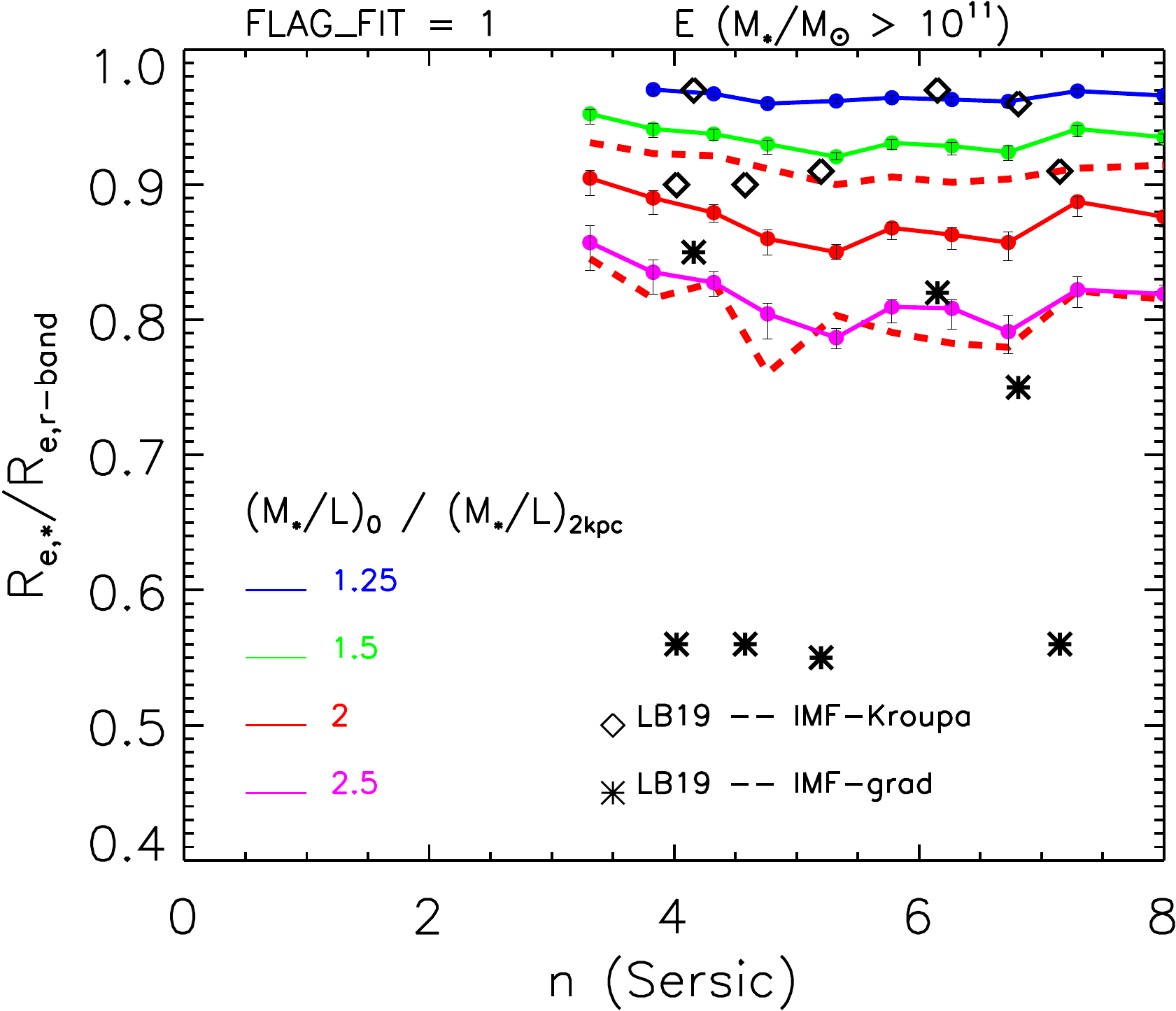} 
  \caption{Left:  Same as Figure~\ref{fig:ratio-n}, except that the gradient runs from the center to 2~kpc, rather than to $0.4R_e$, and curves only show objects with $M_* > 10^{11}M_\odot$, classified as E, and having {\tt FLAG\_FIT = 1}.  In addition, symbols show the size ratio for the seven objects in \citet{LaBarbera2019} when the IMF is assumed to be constant within a galaxy (open diamonds) and when there is a gradient (asterisks).}
  \label{fig:compareLB19-1}
\end{figure}

Figure~\ref{fig:ratio-n} shows the results of this exercise when we set $R_{\rm flat} = 0.4 R_e$ \cite[as suggested by Figure 10 of][]{vD2017} for a number of choices of $\Upsilon_0/\Upsilon_\infty = (M_*/L)_0/(M_*/L)_{0.4R_e}$:  we only show objects which are well fit by a single S{\'e}rsic component, so it makes sense to show $R_{e,*}/R_e$ as a function of $n$.  Clearly, the same $M_*/L$ gradient has a much larger effect if $n$ is large. For $n\ge 5$, $R_{e,*}$ can be smaller than $R_e$ by nearly a factor of 2. This is because of two effects, both of which are related to the fact that profiles with large $n$ are more centrally concentrated.  First, if we think of the integral of $2\pi R\,I(R)\,\Upsilon_\infty$ as giving the total mass without the gradient, then the gradient gives rise to additional mass which is given by integrating $[\Upsilon(R)-\Upsilon_\infty]$ over $2\pi R\,I(R)$.  The ratio of this additional mass to the mass without the gradient is $(\Upsilon_0-\Upsilon_\infty)/\Upsilon_\infty$ times a number which increases as $n$ increases.  Therefore, increasing $\Upsilon_0/\Upsilon_\infty$ will (obviously) result in a larger fractional increase in mass, but the same gradient (i.e. a fixed value of $\Upsilon_0/\Upsilon_\infty$) will produce a larger fractional mass increase if $n$ is larger.  Let us call this extra mass fraction $X_n$, where the subscript is to remind us that it depends on $n$. (E.g., if $\Upsilon_0/\Upsilon_\infty = 2$ and $R_{\rm flat}=R_e/2$, then $X_4\approx 0.17$ whereas $X_8\approx 0.23$.)

Next, we would like to estimate how the new half mass radius -- the scale which encloses $(1+X_n)/2$ -- depends on $X_n$. If the mass from the gradient is confined to scales that are much smaller than the radius which contains half of the rest of the mass, then we can estimate the new $R_{e,*}$ by assuming that it includes all the extra mass $X_n$. The remaining $(1+X_n)/2 - X_n = (1-X_n)/2$ must come from the initial $\Upsilon_\infty$ profile, meaning that the new $R_{e,*}$ is that scale where the original profile contains not $1/2$ but $1/2 - X_n/2$ of the original mass.  Therefore, it is certainly smaller than $R_e$ (even though the total mass is larger).  Since large $n$ profiles are steeper in the central regions, they reach a given mass fraction at a smaller $R/R_e$ than if $n$ is small, and since $X_n$ increases with $n$, the fractional size decrease $R_e/R_{e,*} - 1$ can be significantly larger than the fractional mass increase $X_n$.  In practice, the effect depends on $n$ and how $n$ correlates with stellar mass.  Figure~\ref{fig:alphaSize} shows the results of this exercise.  For example, when $\Upsilon_0/\Upsilon_\infty=2$ then, when the mass increases by 20 percent, the size decreases by more than 30 percent (the size ratio is smaller than 0.7) from what it was originally:  the fractional size change is indeed larger than the fractional change in mass.  This is why we expect accounting for gradients in $M_*/L$ will modify the $R_{e,*}-M_*$ relation (compared to when gradients are weak or are ignored).

\begin{table*}
\centering
M$_*$/L gradients\\
\begin{tabular}{ccccrrrrrrccc}
  \hline
 Bin M$_{r}$ & Bin $\sigma_0$ & N$_{\rm gal}$ & R$_{e,r}$ & $c_0$ & $c_1$ & $c_2$ & $k_0$ & $k_1$ & $k_2$ & r & $c_{\infty}$ & $k_{\infty}$ \\
   & [km s$^{-1}$] & & [kpc] & & & & & & & [kpc] & & \\
\hline
\hline
  E-SR & & & & & &  & & & & & & \\
  \hline

$  -21.5 > \rm{M}_{r} > -22.5$ & $   160 < \sigma_0 < 200$ & $ 57$ & $ 3.55$ & $ 5.300$ & $-0.789$ & $ 0.063$ & $ 3.761$ & $-0.283$ & $ 0.021$ & $6.25$ & $2.81$ & $2.83$ \\
$  -21.5 > \rm{M}_{r} > -22.5$ & $   200 < \sigma_0 < 250$ & $101$ & $ 3.04$ & $ 6.910$ & $-1.537$ & $ 0.158$ & $ 3.776$ & $-0.200$ & $ 0.019$ & $4.75$ & $3.17$ & $3.24$ \\
$  -22.5 > \rm{M}_{r} > -23.5$ & $   200 < \sigma_0 < 250$ & $191$ & $ 6.38$ & $ 7.027$ & $-0.900$ & $ 0.053$ & $ 3.617$ & $-0.099$ & $ 0.006$ & $8.25$ & $3.23$ & $3.19$ \\
$  -22.5 > \rm{M}_{r} > -23.5$ & $   250 < \sigma_0 < 320$ & $181$ & $ 5.62$ & $ 8.426$ & $-1.026$ & $ 0.055$ & $ 3.741$ & $-0.025$ & $-0.002$ & $8.75$ & $3.69$ & $3.34$ \\
$  -23.5 > \rm{M}_{r} > -24.5$ & $   250 < \sigma_0 < 320$ & $ 51$ & $11.80$ & $ 8.351$ & $-0.931$ & $ 0.046$ & $ 3.916$ & $-0.118$ & $ 0.008$ & $8.75$ & $3.72$ & $3.47$ \\

 \hline
  E-FR & & & & & & & & & & & & \\
  \hline

$  -20.5 > \rm{M}_{r} > -21.5$ & $   160 < \sigma_0 < 200$ & $ 59$ & $ 1.55$ & $ 6.616$ & $-2.121$ & $ 0.292$ & $ 3.123$ & $-0.021$ & $-0.014$ & $3.75$ & $2.78$ & $2.77$ \\
$  -21.5 > \rm{M}_{r} > -22.5$ & $   160 < \sigma_0 < 200$ & $ 85$ & $ 2.43$ & $ 5.859$ & $-1.551$ & $ 0.154$ & $ 2.659$ & $-0.150$ & $ 0.001$ & $4.75$ & $1.98$ & $1.97$ \\
$  -21.5 > \rm{M}_{r} > -22.5$ & $   200 < \sigma_0 < 250$ & $137$ & $ 2.71$ & $ 7.537$ & $-1.535$ & $ 0.122$ & $ 3.287$ & $-0.084$ & $-0.001$ & $5.25$ & $2.84$ & $2.79$ \\
$  -22.5 > \rm{M}_{r} > -23.5$ & $   200 < \sigma_0 < 250$ & $ 79$ & $ 4.31$ & $ 6.392$ & $-0.850$ & $ 0.047$ & $ 2.881$ & $-0.068$ & $ 0.003$ & $8.75$ & $2.54$ & $2.46$ \\
$  -22.5 > \rm{M}_{r} > -23.5$ & $   250 < \sigma_0 < 320$ & $ 78$ & $ 5.19$ & $ 9.354$ & $-1.186$ & $ 0.061$ & $ 3.479$ & $-0.064$ & $ 0.006$ & $8.75$ & $3.64$ & $3.37$ \\

 \hline
  S0 & & & & & & & & & & & & \\
 \hline

$  -19.5 > \rm{M}_{r} > -20.5$ & $   100 < \sigma_0 < 125$ & $ 51$ & $ 0.90$ & $ 4.820$ & $-1.803$ & $ 0.344$ & $ 2.732$ & $ 0.108$ & $-0.096$ & $2.25$ & $2.50$ & $2.49$ \\
$  -19.5 > \rm{M}_{r} > -20.5$ & $   125 < \sigma_0 < 160$ & $ 44$ & $ 0.85$ & $ 5.552$ & $-1.351$ & $ 0.091$ & $ 3.349$ & $-0.182$ & $ 0.027$ & $2.25$ & $2.97$ & $3.08$ \\
$  -20.5 > \rm{M}_{r} > -21.5$ & $   125 < \sigma_0 < 160$ & $ 72$ & $ 1.33$ & $ 3.749$ & $-1.108$ & $ 0.177$ & $ 2.694$ & $-0.399$ & $ 0.057$ & $3.25$ & $2.02$ & $1.93$ \\
$  -20.5 > \rm{M}_{r} > -21.5$ & $   160 < \sigma_0 < 200$ & $ 82$ & $ 1.42$ & $ 6.746$ & $-1.671$ & $ 0.179$ & $ 3.260$ & $-0.104$ & $-0.004$ & $4.25$ & $2.87$ & $2.74$ \\
$  -21.5 > \rm{M}_{r} > -22.5$ & $   160 < \sigma_0 < 200$ & $ 71$ & $ 2.38$ & $ 5.291$ & $-0.955$ & $ 0.063$ & $ 2.732$ & $-0.277$ & $ 0.020$ & $4.75$ & $2.18$ & $1.87$ \\
$  -21.5 > \rm{M}_{r} > -22.5$ & $   200 < \sigma_0 < 250$ & $ 42$ & $ 2.67$ & $ 7.209$ & $-0.952$ & $ 0.047$ & $ 3.285$ & $ 0.052$ & $-0.030$ & $5.25$ & $3.50$ & $2.72$ \\

 \hline
 \end{tabular}
\caption{Coefficients of quadratic fits to the $M_*/L$ profiles shown in Figures~\ref{fig:profileE}, that are associated with IMF-gradients ($c_i$) or a fixed (Kroupa) IMF ($k_i$), and were estimated from spectra that were stacked in bins of $r$-band luminosity (column 1), central velocity dispersion (column 2), and morphology. Column 3 reports the number of galaxies in the bin. Column 4 gives the median value of the half-light radius for the galaxies in the bin. Columns 5-7 give the $c_i$; columns 8-10 give the $k_i$; column 11 gives the scale beyond which (IMF-gradient) $M_*/L$ is constant; columns 12 and 13 give the value of this constant for a variable or Kroupa IMF. }
\label{tab:Cat}
\end{table*}

Figure~\ref{fig:ratio-BT} shows a similar analysis of the objects which are better described by two-components.  Therefore, it shows $R_{e,*}/R_e$ as a function of B/T rather than $n$ (there are no objects with B/T$>0.7$ because we only show bins containing at least 25 objects). Although the effects are qualitatively similar -- increasing the $M_*/L$ gradient reduces the half-mass radius -- for these objects too, accounting for $M_*/L$ gradients will modify the $R_{e,*}-M_*$ relation. However, the dependence on B/T is weaker than that on $n$ shown in Figure~\ref{fig:ratio-n} and the scatter is larger.  The larger scatter arises as follows.
  Let $n_b$ and $R_b$ denote the Sersic index and half-light radius of the inner `bulge' component, and suppose for now that the gradient only extends to scales that are smaller than $R_b$.  Then, one might have thought we would see a similar (tight) correlation to that shown in Figure~\ref{fig:ratio-n} if we had plotted $R_{b,*}/R_b$ versus $n_b$.  However, the gradient extends to $0.4R_e = 0.4R_b\, (R_e/R_b)$, so the fact that $R_e/R_b$ can vary between objects will introduce scatter in the effect of the gradient on $R_{b,*}/R_b$.  This ratio will also introduce scatter in the y-axis of Figure~\ref{fig:ratio-BT}, which shows $R_{e,*}/R_e$ rather than $R_{b,*}/R_b$.  Finally, the x-axis  of Figure~\ref{fig:ratio-BT} shows B/T rather than $n_b$, and this produces further scatter. 

\begin{figure*}
  \centering
  \includegraphics[width=0.85\linewidth]{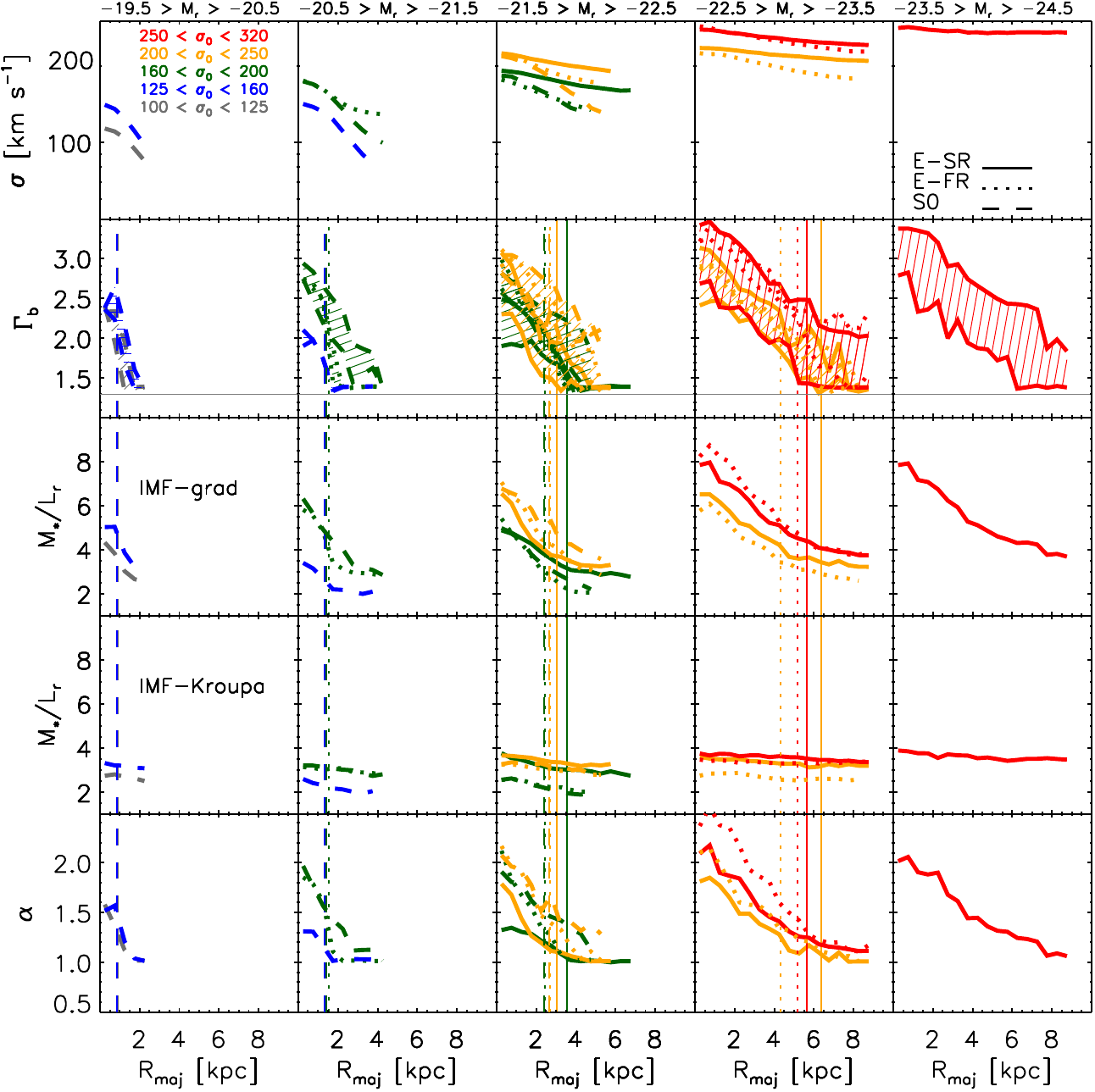}
  \caption{Profiles of velocity dispersion (top row), bimodal IMF parameter $\Gamma_b$ which results from fitting the observed spectral features (second from top row), associated $M_*/L$ ratio (middle row), $M_*/L$ ratio when the IMF is forced to be Kroupa on all scales (second from bottom row), and ratio of the two $M_*/L$ values ($\alpha \equiv M_{*\rm IMF-grad}/M_{*\rm IMF-Kroupa}$ -- bottom row) for a number of bins in luminosity (which increases from left to right) and central velocity dispersion (different colors, as labeled). Quadratic fits to the `IMF-grad' and `IMF-Kroupa' trends are reported in Table~\ref{tab:Cat}.  Solid, dotted and dashed curves show slow rotating and fast rotating ellipticals, and S0s.  Vertical lines in each panel show $R_e/2$ for each sample. Bottom panel shows that IMF variations produce much stronger $M_*/L$ gradients. Shaded regions in the second from top row show a crude estimate of the systematic uncertainties in determining $\Gamma_b$ which arise from the correction for emission in the H$_\beta$ line (see text for details).}
  \label{fig:profileE}
\end{figure*}

We now consider the case in which gradients extend to a fixed physical scale \cite[as suggested by Figure 5 of][]{LaBarbera2019}, rather than a fixed fraction of $R_e$.  In this case, gradients result in a smaller fractional mass increase for two reasons:  First, $R_e$ is larger for massive objects, so 2~kpc is a smaller fraction of $R_e$ at large $M_*$, meaning that gradients result in a smaller $X_n$ for massive galaxies.  In addition, massive galaxies typically have the largest values of $n$ (we show no results at $n<3.7$ because there are fewer than 25 galaxies per bin at smaller $n$), for which the size change is otherwise most dramatic.  So limiting the gradient to 2~kpc reduces the changes associated with large $n$.  The curves in Figure~\ref{fig:compareLB19-1} show the ratio of sizes for objects with $M_* > 10^{11}M_\odot$, classified as E, and having {\tt FLAG\_FIT = 1}: notice that the trends are much reduced with respect to Figure~\ref{fig:ratio-n}.  They are also noisier, because 2 kpc is such a small and variable fraction of the galaxy size.

As a simple consistency check of our methodology, we have taken the seven massive galaxies studied by \citet{LaBarbera2019}.  For these objects, they provide stellar population parameters when the IMF is assumed to be constant within a galaxy and when there is a gradient as reported in their Figure 5.  We have used their determinations to estimate $R_{e,*}$ and $M_*$ for each object.  Asterisks and open diamonds show the estimates with and without a gradient; clearly, accounting for gradients significantly reduces the sizes.  For four of their objects, this effect is dramatic:  $R_{e,*}/R_e \sim 0.55$.  This is because these four objects have $R_e\sim 2$~kpc, so the gradient, which is restricted to scales smaller than 2~kpc, has a big effect.  The other three objects have much larger $R_e$, so, for them, the effect of the gradient is much smaller.

We end with the observation that, since $R_{e,*}/R_e$ depends on both $n$ and the $M_*/L$ gradient, this ratio will evolve if either or both evolve.  We discuss this further in Section~\ref{sec:hiz}.

\subsection{$M_*/L$ gradients in MaNGA}
As described in Section~\ref{stacks} (see \citealt{Bernardi2022} for details), we use the MILES+Padova SSP models to estimate age, [M/H], [$\alpha$/Fe] and IMF profiles. These SSP parameters can then be turned into profiles of $M_*/L_r$ to determine the $R_{e,*}-M_*$ relation. Briefly, a given age and [M/H] define a turn-off mass (the MILES models do not yet include a dependence of the turn-off mass on [$\alpha$/Fe]).  When combined with $\Gamma_b$, the turn-off mass defines the mass in stars still on the main sequence, $M_{\rm s}$, and the mass in remnants (white dwarfs, neutron stars, black holes), $M_{\rm r}$.  The quantity called $M_*$ is the sum of these two: $M_{\rm s} + M_{\rm r}$.   I.e., as is common practice \cite[e.g.][]{Vazdekis2015}, $M_*$ does {\em not} include the mass in gas.  The four SSP parameters also define a spectrum.  To get $L_r$, we simply integrate over this spectrum using the $r$-band filter.  We then combine $M_*$ and $L_r$ to get $M_*/L_r$ as a function of (age, [M/H], [$\alpha$/Fe], $\Gamma_b$).  In practice, the MILES spectra are provided in a few discrete bins in SSP parameter values, and we interpolate to get $M_*/L_r$ at the values we want.  We include non-zero $[\alpha$/Fe] to get the MILES+Padova values by scaling the BaSTI $L_r$ values, similarly to how \cite{DS2019} scale the absorption line strengths.

The top panels of Figure~\ref{fig:profileE} show the average velocity dispersion profiles in each bin of luminosity and velocity dispersion as labeled: solid, dotted and dashed curves show $\sigma(R)$ for slow and fast rotating ellipticals, and S0s. In what follows, we present results based on stacking spaxels in $R$ (our results are very similar when we do scale by $R_e$ before stacking). Profiles are shown out to 8~kpc or $\sim 1.5$ $R_e$ for smaller galaxies (remember that $R_e$ is the half-light radius of the truncated profile along the semimajor axis).

The next set of panels shows $\Gamma_b(R)$, the parameter which describes the shape of the bimodal IMF: recall that a Kroupa IMF has $\Gamma_b = 1.3$ and larger values of $\Gamma_b$ are more bottom heavy. Clearly, $\Gamma_b$ is large in the central regions, and decreases outwards.  Vertical lines show $R_e/2$ (i.e. half of the half light radius):  although not the main focus of our study, it is worth noting that, at fixed luminosity, the E-SRs with large $\sigma_0$ have smaller $R_e$, in qualitative agreement with the virial theorem (if luminosity is approximately proportional to mass).

The shaded regions show a crude estimate of the systematic uncertainties in determining $\Gamma_b$ which arise from the correction for emission in the H$_\beta$ line (see \citealt{Bernardi2022} for more discussion of why this is the dominant systematic in the measurements).  
Typically, $\Gamma_b$ reaches Kroupa (solid horizontal line close to the bottom of each panel) a little beyond $R_e/2$.  Evidently a model in which gradients scale with $R_e$ is more realistic than one in which they are confined to a fixed physical scale. (This remains true if we stack in $R/R_e$ rather than $R$ in kpc.)
In addition, at fixed $L$, $\Gamma_b$ is larger if $\sigma_0$ is larger, but this is only obvious if one compares objects of the same morphological type.  We have also measured gradients in age, metallicity and element abundances -- e.g., we find a strong correlation between $\Gamma_b$ and metallicity which is qualitatively consistent with recent work reporting more bottom-heavy IMFs at higher metallicities \citep{Liang2021}, and that metallicity determines color but IMF determines $M_*/L$ in old galaxies -- but we discuss these trends elsewhere \citep{Bernardi2022}.

Here we are mainly interested in the associated $M_*/L$ profiles, which
depend on the SSP estimated age, [M/H], [$\alpha$/Fe] and $\Gamma_b$ values. The $M_*/L$ profiles are shown in the next set of panels. (These $M_*/L$ values are for the SDSS $r-$band, so they are in units of $M_\odot/L_{\odot,r}$.) For each $R$, the lines show the average of the $M_*/L$ values associated with the upper and lower limits of $\Gamma_b$. Quadratic fits to these trends are reported in Table~\ref{tab:Cat}.

\begin{figure}
  \centering
  \includegraphics[width=0.85\linewidth]{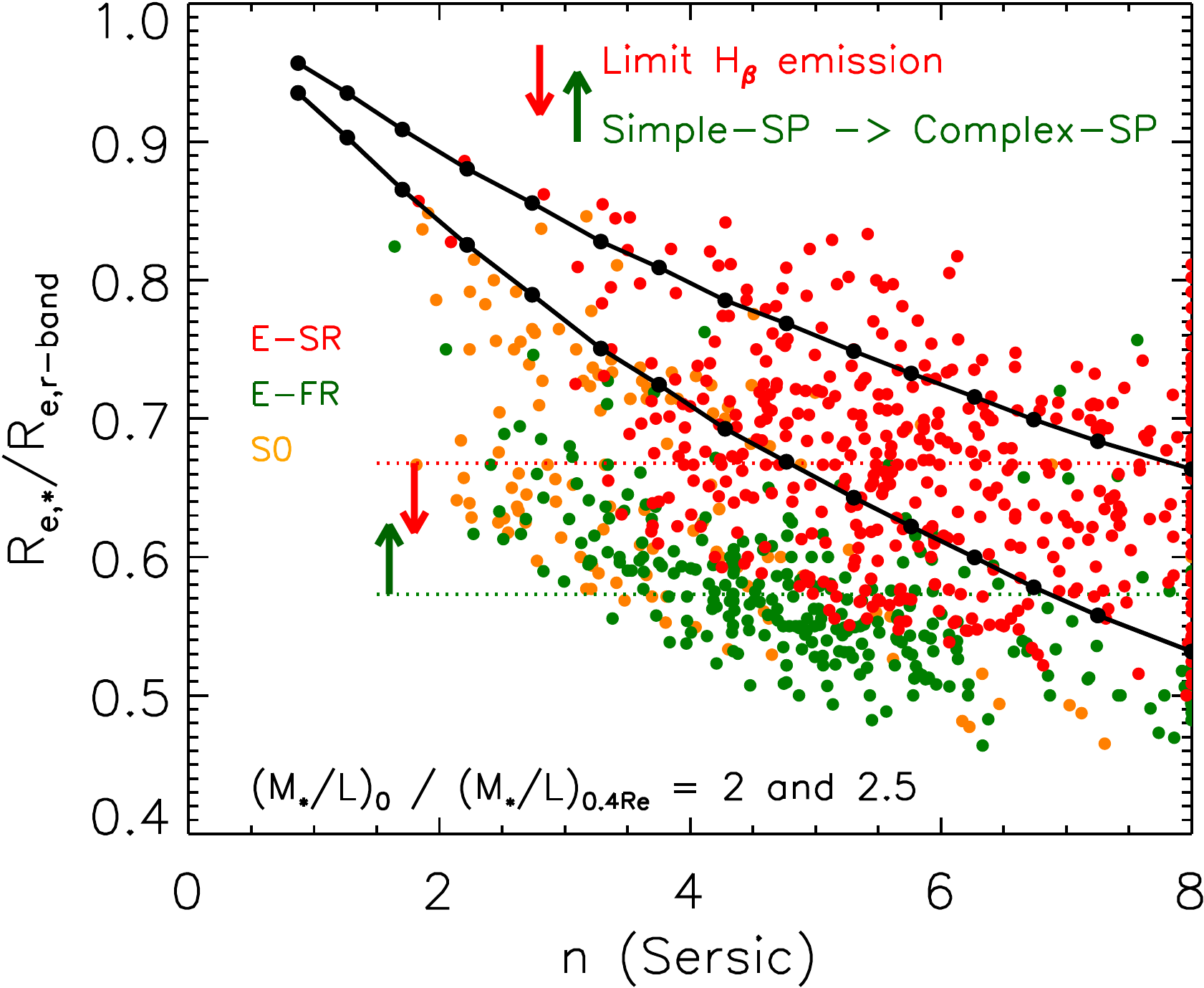}
  \includegraphics[width=0.85\linewidth]{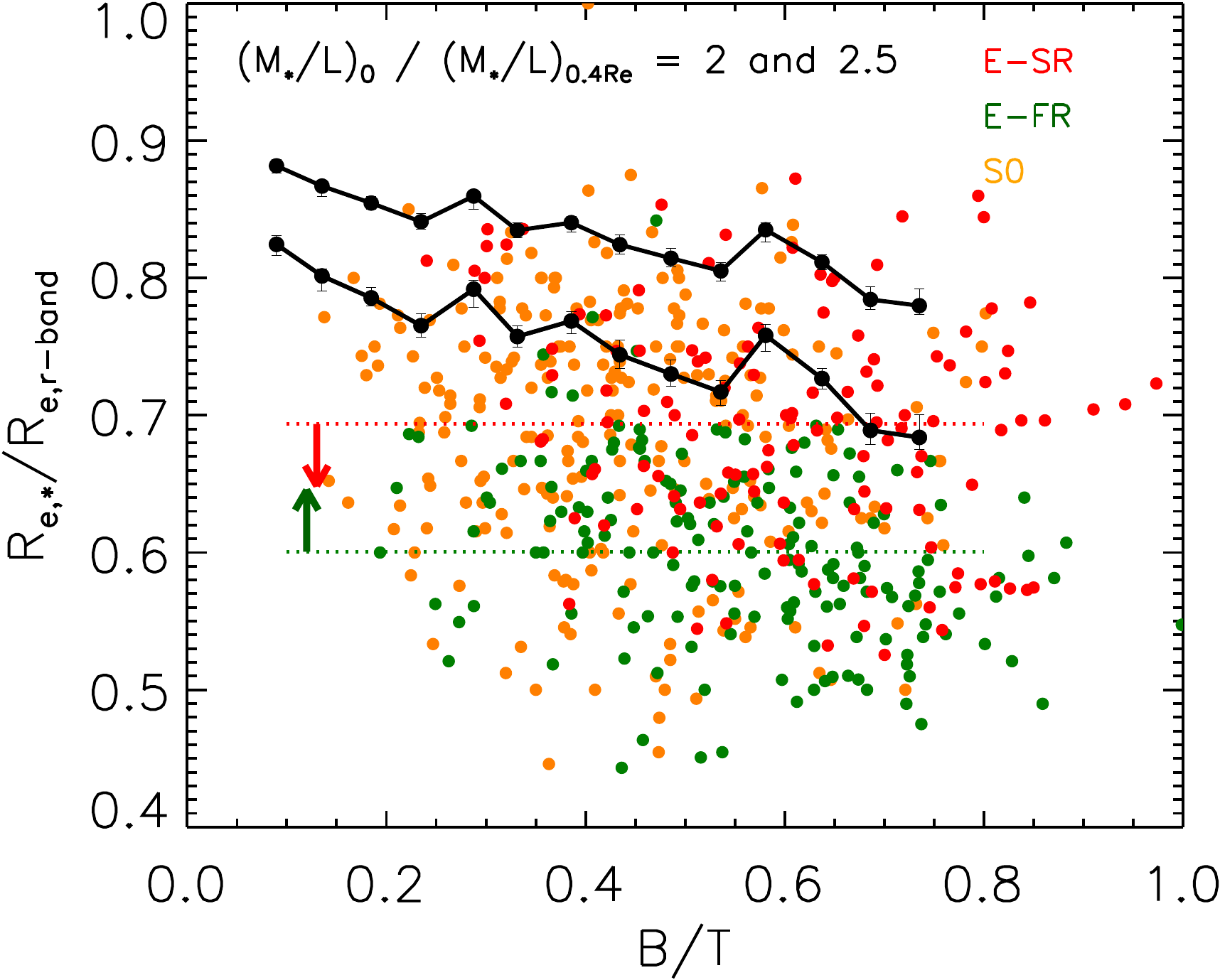} 
  \caption{Dependence of $R_{e,*}/R_e$ on $n$ (top) and B/T (bottom) for the IMF-driven $M_*/L$ gradients shown in the middle row of Figures~\ref{fig:profileE}, for all three morphological types. Thin dotted lines show the median values of E-SR and E-FRs. Red and green arrows in top panel show how systematics might shift the E-SR and E-FR results.
    Solid black curves in the two panels show the bottom two relations from Figures~\ref{fig:ratio-n} and~\ref{fig:ratio-BT}, which assume that equation~(\ref{eq:toy}) describes the gradients.  The IMF-driven $M_*/L$ gradients in MaNGA can reduce the sizes by nearly a factor of two, especially at large $n$.}
  \label{fig:mangaBTn}
\end{figure}

Notice that $M_*/L$ tends to be large in the central regions and decrease outwards, reaching the associated Kroupa values just beyond $R_e/2$.  The most luminous objects with the largest $\sigma_0$ tend to have the largest $M_*/L$ values on all scales.  The panels, which are second from bottom row, show the $M_*/L$ profiles if we fix the IMF to Kroupa:  the dependence on $L$ and $\sigma_0$ is qualitatively similar, but the $M_*/L$ gradients are much less pronounced.\footnote{We note in passing that the fact that they are small is consistent with the small effects seen by \cite{Tortora2011} and \cite{califa2019} which were based on color-gradients and which, as we have noted, cannot account for IMF gradients.}

\begin{figure}
  \centering
  \includegraphics[width=0.825\linewidth]{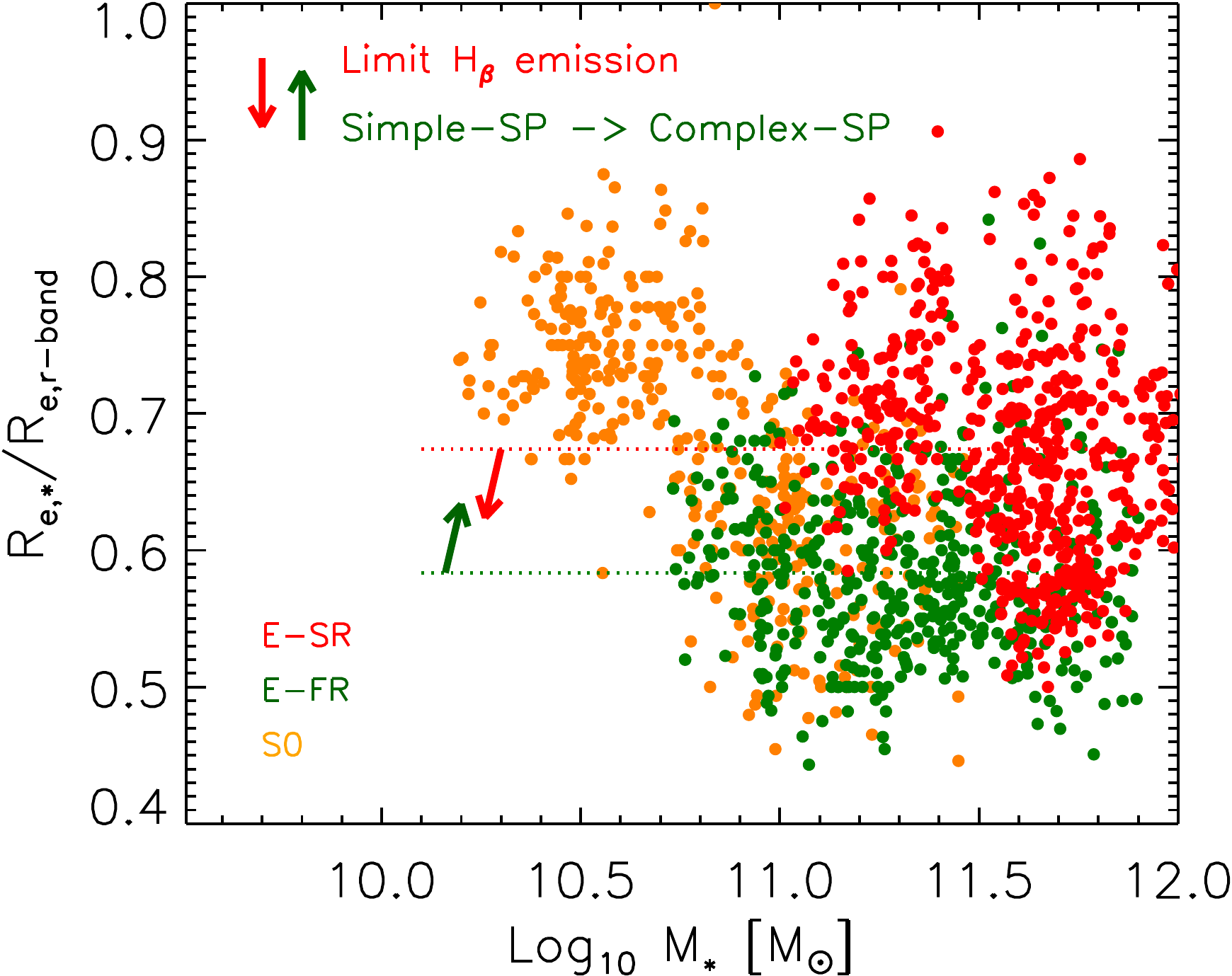}  
  \caption{Dependence of $R_{e,*}/R_e$ on $M_*$, where both $R_*$ and $M_*$ were estimated from the IMF-driven $M_*/L$ gradients shown in the middle row of Figures~\ref{fig:profileE}, for all three morphological types. Red and green arrows show potential systematic corrections to the E-SR and E-FR values. The IMF-driven $M_*/L$ gradients in MaNGA can reduce the sizes by nearly a factor of two, especially at the highest masses.  }
  \label{fig:mangaM*}
\end{figure}
\begin{figure}
  \centering
  \includegraphics[width=0.825\linewidth]{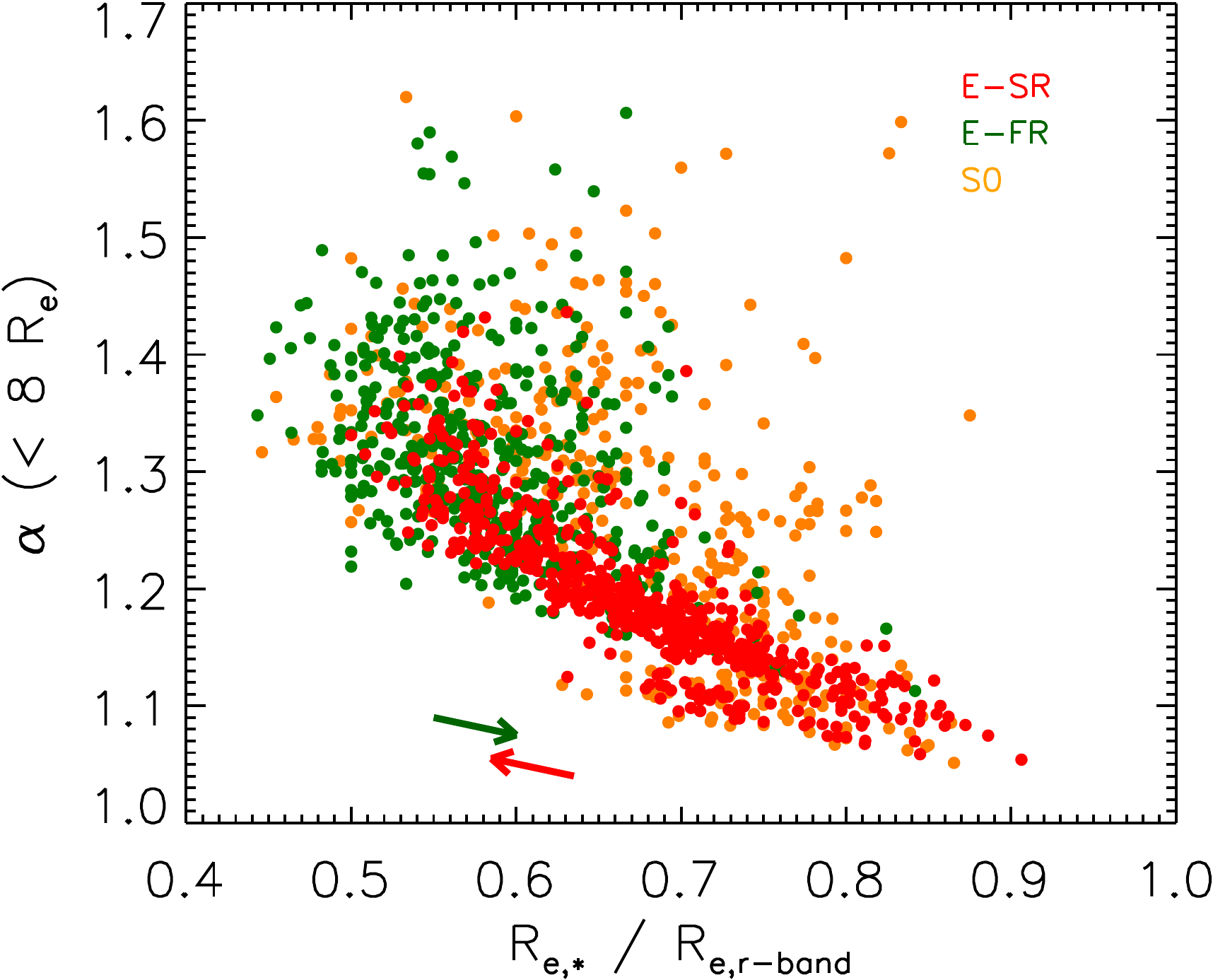}
  \includegraphics[width=0.825\linewidth]{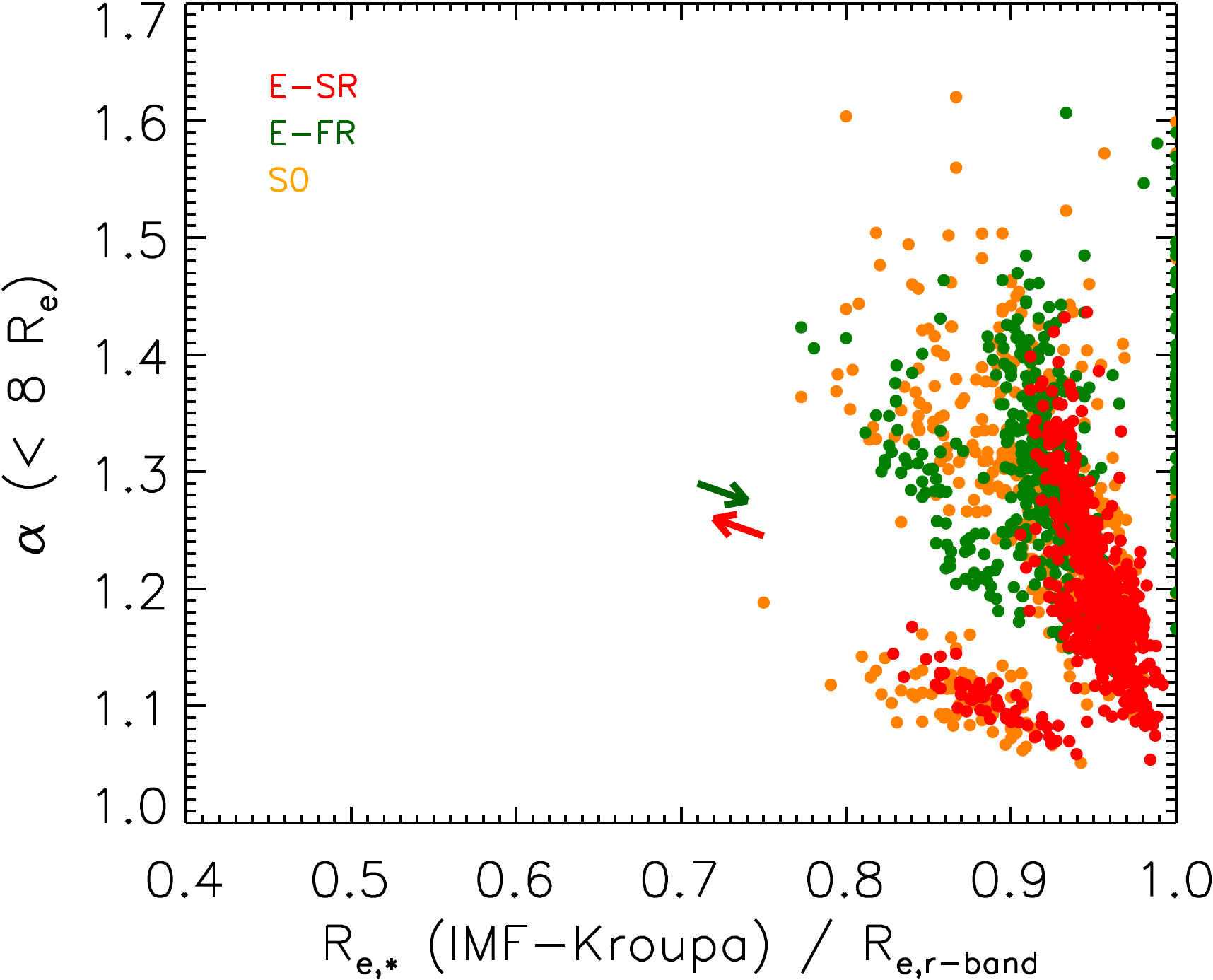}
  \caption{Top: Correlation between $R_{e,*}/R_e$ and the IMF mismatch parameter $\alpha\equiv M_{*\rm IMF-grad}/M_{*\rm IMF-Kroupa}$ (i.e. when the fits provided in Table~\ref{tab:Cat} are inserted in equation~\ref{eq:totM*}, and the integral is computed out to $8R_e$) for the three morphological types.  Red and green arrows show potential systematic corrections to the E-SR and E-FR values.
    Bottom:  Same as top, except that now $R_{e,*}$ is that for a Kroupa IMF.  Ellipticals in this panel tend to have size ratios of order unity, but in the top panel the size ratios are smaller, indicating that most of the change in mass and size is due to the IMF gradients. }
  \label{fig:mangaA}
\end{figure}

\begin{figure*}
  \centering
  \includegraphics[width=0.9\linewidth]{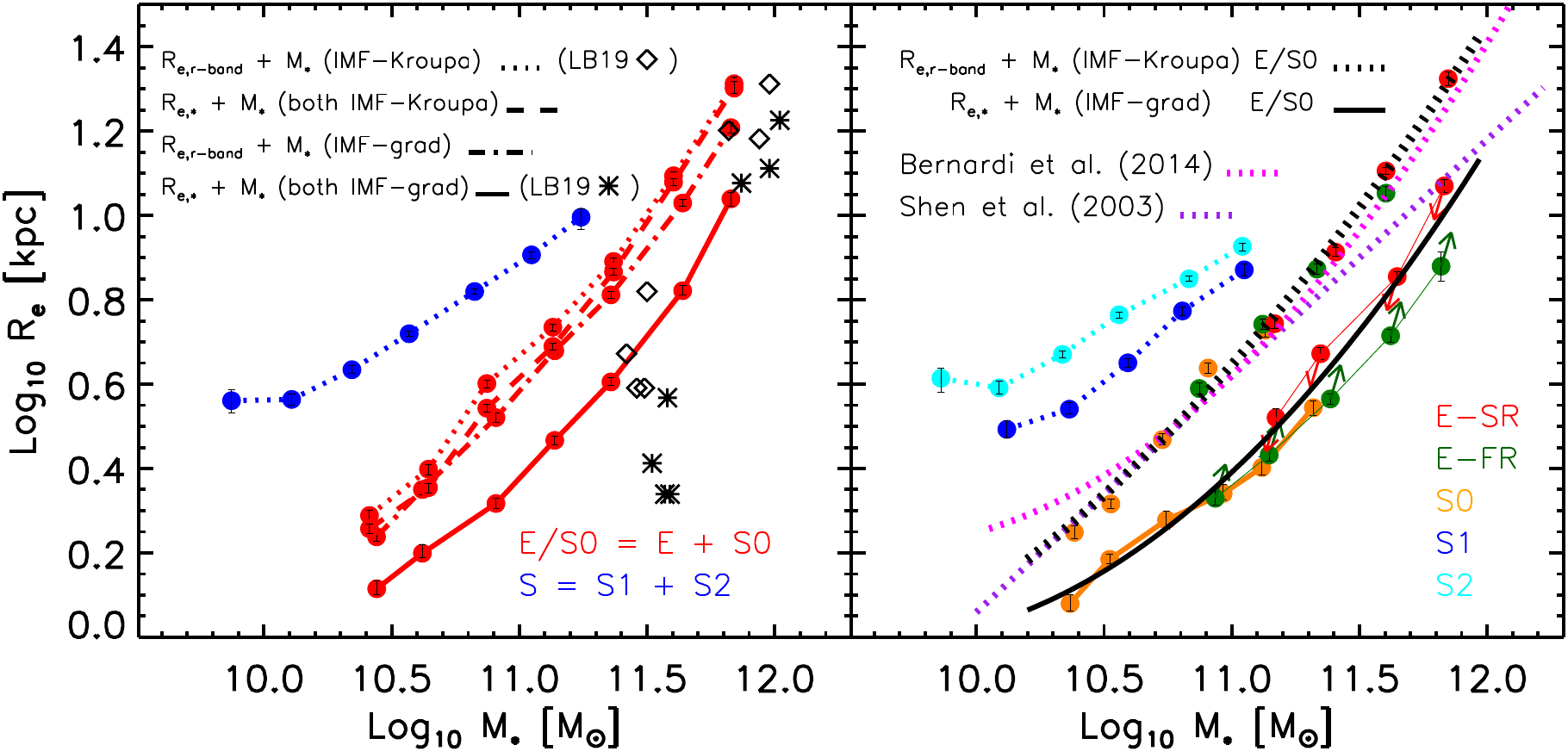}
  \caption{Left:  Dependence of size-mass relation on $M_*/L$ gradients for early-type galaxies (red).  Blue curve shows this relation for later-types (blue).  Dotted curves show half-light radius $R_e$ versus $M_*$ which result from assuming the IMF is fixed to Kroupa; dashed curves use the same (fixed IMF) $M_*$ but the corresponding $R_{e,*}$ (size change as in the bottom panel of Figure~\ref{fig:mangaA}); dot-dashed curve shows $R_e$ versus the $M_*$ which includes the IMF-driven gradient, and solid curve uses the associated $R_{e,*}$ (size change as in the top panel of Figure~\ref{fig:mangaA}).  For early-types, this final, self-consistent $R_{e,*}-M_*$ relation is offset to smaller sizes by nearly 0.3~dex (at $M_*\sim 10^{11}M_\odot$) compared to the original $R_e-M_*$ relation.  Symbols show estimated sizes and masses of the seven $z\sim 0$ galaxies from La Barbera et al. (2019), which we showed in Figure~\ref{fig:compareLB19-1}.
    Right: Comparison of $R_e-M_*$ relations when the IMF is fixed to Kroupa, and $R_{e,*}-M_*$ when IMF-driven gradients are included, for the different morphological types (as indicated). Red and green arrows show potential systematic corrections to the E-SR and E-FR values. Smooth black dotted and solid curves show fits to the relations defined by early-type galaxies, the parameters of which are reported in Table~\ref{tab:RMfits}; pink and purple dotted curves show fits to $R_e-M_*$(Kroupa) from the literature.}
  \label{fig:MsRe}
\end{figure*}

The bottom set of panels highlights this:  the quantity $\alpha$ shown is the ratio of the $M_*/L$ curves in the two panels above (i.e. the curves labeled `IMF-grad' divided by those labeled `IMF-Kroupa').  Notice that $\alpha > 2$ in the central regions of the most luminous galaxies.  This illustrates graphically why we refer to the $M_*/L$ gradients (shown in the panels labeled `IMF-grad') as being IMF{\em -driven}.  In some cases, the IMF-driven gradients have a central value that is more than $2\times$ greater than the asymptotic value at large $R$.

Having studied the mass ratios within $R_e$, Figure~\ref{fig:mangaBTn} shows the size ratios $R_{e,*}/R_e$ as a function of S{\'e}rsic index $n$ (for objects with {\tt FLAG\_FIT=1}) and B/T (objects with {\tt FLAG\_FIT=2} or {\tt 0}). In all cases, we determine $R_{e,*}$ by using the quadratic fits to $M_*/L$ from Table~\ref{tab:Cat}, for the relevant stack, as $\Upsilon(R)$ in equation~(\ref{eq:R*}). The red and green arrows in the top panel show potential systematic corrections to the E-SR and E-FR values.  As \cite{Bernardi2022} discuss, for E-SRs, these arise because the emission is very weak, so correcting for it is quite uncertain.  For E-FRs, the emission correction is less uncertain, but the younger ages and smaller [$\alpha$/Fe] values suggest that stellar population parameters determined by fitting an SSP to the measurements may result in $M_*/L$ values that are slightly biased because the population may, in fact, be more complex.  While these systematics may matter in detail, they do not affect the general conclusion that $R_{e,*}/R_e$ may be substantially smaller than unity.

The size ratios in Figure~\ref{fig:mangaBTn} are consistent with expectations from the simple model of equation~(\ref{eq:toy}):  solid black curves show the bottom two curves of Figures~\ref{fig:ratio-n} and~\ref{fig:ratio-BT}.  (The agreement is better if we set $R_{\rm flat}=0.6R_e$, rather than $0.4R_e$, in equation~\ref{eq:toy}).
Evidently, the $M_*/L$ gradients in MaNGA produce effects that are comparable to or stronger than those from equation~(\ref{eq:toy}): IMF-driven gradients in MaNGA can reduce the sizes by as much as a factor of two, especially at the largest $n$ or at the largest masses (see Figure~\ref{fig:mangaM*}).
The `pearls on a string' like effects in this and other figures arise because there is a range of $n$ values in each $L$ and $\sigma_0$ bin, and $R_{e,*}/R_e$ is correlated tightly with $n$ (see Figure~\ref{fig:ratio-n}).  In contrast, $R_{e,*}/R_e$ at fixed B/T has much larger scatter (Figure~\ref{fig:ratio-BT}) so the string of pearls defining a sharp lower limit is much less evident.

It is worth making one other point in this context.  Without the black curves to guide the eye, the top panel in Figure~\ref{fig:mangaBTn} appears to show little correlation between $R_{e,*}/R_e$ and $n$.  This is in agreement with the rightmost panels of Figure~7 in \cite{Szomoru2013}, who conclude that ``the difference between half-mass size and half-light size correlates very weakly with galaxy structure''.  Our results show clearly that this is {\em not} because galaxy structure does not matter -- the toy model shows that $n$ matters very much!  What happens is that the tight correlation with $n$ is masked by the large scatter in the strength of the $M_*/L$ gradient.

The analog of Figure~\ref{fig:alphaSize} is shown in Figure~\ref{fig:mangaA}, which shows how $\alpha\equiv M_{*\rm IMF-grad}/M_{*\rm IMF-Kroupa}$, the IMF `mismatch' parameter (i.e. when the fits provided in Table~1 are inserted in equation~1, and the integral is computed out to $8R_e$) correlates with $R_{e,*}/R_e$.  The top panel shows that, typically, the largest mass increases (relative to Kroupa) are associated with the largest size decreases (relative to the half-light radius). 
These trends are rather similar to the toy model expectations.  The bottom panel shows that the size ratio when the IMF is fixed to Kroupa is close to unity.  This shows explicitly that most of the size change in the top panel is due to the IMF gradient.

Note that Figures~\ref{fig:mangaBTn}--\ref{fig:mangaA} only show results for the objects which contributed to our stacks (see Table~\ref{tab:Cat}): remember that we only use bins (in luminosity and $\sigma_0$) which include at least 40 galaxies to avoid cases in which a few objects dominate the stack (Section~\ref{stacks}).

\subsection{The size-mass correlation in MaNGA}
We are finally ready to consider the effect of gradients on the size-mass correlation.  
Filled circles connected by dotted line in the left hand panel of Figure~\ref{fig:MsRe} show median half-light radius $R_e$ versus median $M_*$ for a Kroupa IMF (small error bars show the (Poisson) error on the median size in the bin), and circles connected by dashed curves replace $R_e\to R_{e,*}$ but use the same Kroupa IMF to compute $R_{e,*}$ and $M_*$ (i.e., the size change as in the bottom panel of Figure~\ref{fig:mangaA}).  The two relations are quite similar but this is not surprising because, at fixed IMF, gradients are small.  In contrast, the dot-dashed curve shows $R_e$ versus the $M_*$ which includes the IMF-driven gradient, and solid curve uses the associated $R_{e,*}$ (size change as in the top panel of Figure~\ref{fig:mangaA}).  Whereas the $R_e-M_*$ relation is not too different from the others -- recall that the increase in $M_*$ is not too dramatic -- the final, self-consistent $R_{e,*}-M_*$ relation is offset to smaller sizes by nearly 0.3~dex (at $M_*\sim 10^{11}M_\odot$) compared to the original $R_e-M_*$ relation.

The right hand panel compares the $R_e-M_*$ relations when the IMF is fixed to Kroupa, and the $R_{e,*}-M_*$ relations when IMF-driven gradients are included, for the different morphological types (as indicated). We only show large symbols if there are at least 25 objects in the bin. Whereas the corresponding $R_e-L_r$ relations show that E-FRs are slightly smaller than E-SRs of the same $L_r$ \citep{Bernardi2019}, these differences are reduced if the IMF is fixed when converting from $L$ to $M_*$ (dotted), but not if IMF-driven gradients are allowed (thin solid).  However, there could be systematic corrections. The red and green arrows show potential systematic corrections to the E-SR and E-FR values (see Figure~\ref{fig:mangaBTn} and related discussion). Smooth black dotted and solid curves show fits to these relations defined by early-type galaxies which we report in Table~\ref{tab:RMfits} (we fit to the objects themselves, not to the large symbols).

\begin{table}
    \centering
    \begin{tabular}{cc|rcc}
    \hline
    Relation & IMF & $a_0$ & $a_1$ & $a_2$ \\
    \hline
    \hline
    $R_{e,*}-M_{*}$ & IMF-grad & 0.0230 & $0.1666$ & 0.2014\\  
    $R_e-M_{*}$  & Kroupa & 0.0941 & $0.4224$  & 0.1285\\  
    \hline
    \end{tabular}
    \caption{Coefficients of quadratic fits to the MaNGA size-mass relations (the black solid and dotted curves in the right hand panel of Figure~\ref{fig:MsRe}), $\log(R/{\rm kpc}) = a_0 + a_1 m + a_2 m^2$ where $m\equiv \log(M/10^{10}M_\odot)$, when IMF-driven gradients are accounted for (top; $R_{e,*}$ is the half-mass radius) and when they are ignored (bottom; $R_e$ is the half-light radius).}
    \label{tab:RMfits}
\end{table}

\begin{figure*}
    \centering
    \includegraphics[width=0.85\textwidth]{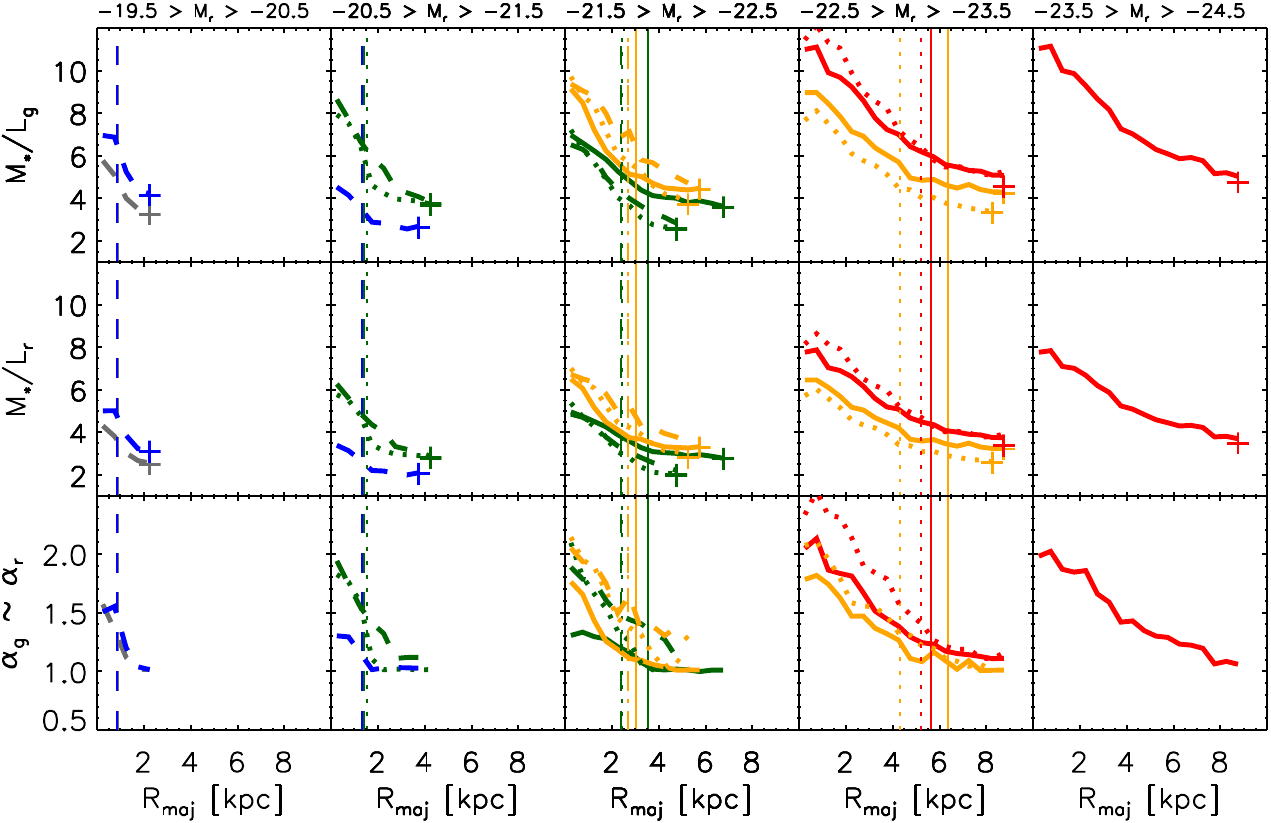}
    \caption{Stellar mass to light ratios in the $g$- and $r$-bands, when the IMF is allowed to vary with scale, for the same $L$ (panels), $\sigma_0$ (colors) and morphological types (line type) as Figure~\ref{fig:profileE}.  Cross indicates the large scale value when the IMF is fixed to Kroupa.  Bottom panel shows the ratio of each curve to that obtained when the IMF is fixed to Kroupa.  Although $M_*/L_g > M_*/L_r$, this is also true for Kroupa, so the ratio $\alpha_g\sim\alpha_r$.  }
    \label{fig:alphagr}
\end{figure*}

To establish that the difference between $R_e-M_*$ and $R_{e,*}-M_*$ relations is not due to problems with the original (fixed IMF) $R_e-M_*$ relation in MaNGA, the dotted pink and purple curves show previous estimates of this same relation at low $z$ in the SDSS main galaxy sample (stellar masses were scaled to Kroupa IMF using the values provided in Table~2 of \citealt{Bernardi10}).  Whereas \cite{Shen2003} fit a simple power law to the $R_e-M_*$ relation, \cite{Bernardi2014} noted there was curvature towards larger sizes at the highest masses, which their fit allowed for.  The $R_e-M_*$ relation we see in MaNGA shows similar curvature.  Compared to any of these fits, the $R_{e,*}-M_*$ relation in MaNGA is clearly offset to smaller sizes.

Finally, as another consistency check, the symbols in the panel on left show the $R_e-M_*$ (open diamonds) and $R_{e,*}-M_*$ (asterisks) correlations for the seven objects studied by LB19. (Each asterisk sits down and slightly to the right of its diamond.)  While they tend to sit slightly below our relations, the relative differences are similar.

\subsection{Results in the SDSS $g$-band}
In the next section, we compare the $z\sim 0$ MaNGA size-mass correlations with those estimated at higher redshifts.  The high-$z$ sizes were corrected to restframe 5000~\AA.  Since this lies between the SDSS $g$- and $r$-bands, we have repeated all the $r$-band analyses of the previous section, but now in the $g$-band instead.  In practice, the MILES+Padova SSP parameters which we used to obtain $M_*/L_r$ also predict $M_*/L_g$ (see \citealt{Bernardi2022} for details).  So, we used these to produce the profiles shown in the top panels of Figure~\ref{fig:alphagr}.  These $M_*/L_g$ profiles are for an IMF which varies with scale, and the middle set of panels show the corresponding $M_*/L_r$ profiles (from Figure~\ref{fig:profileE}).

Notice that $M_*/L_g > M_*/L_r$ in all cases.  However, in all cases, this is also true when the IMF is fixed to Kroupa.  (To avoid clutter, a cross marks the Kroupa value of $M_*/L$ on large scales; the panels which are second from bottom in Figure~\ref{fig:profileE} show that $M_*/L_r$ is nearly constant, and this is true for $M_*/L_g$ as well.)  Indeed, the bottom panels show $\alpha_g$, the ratio of these $M_*/L_g$ profiles to those when the IMF is fixed to Kroupa.  Comparison with the $\alpha_r$ profiles shown in the bottom panel of Figure~\ref{fig:profileE} shows that, to a very good approximation, $\alpha_g\approx \alpha_r$.  To understand why, note that $\alpha_r/\alpha_g$ is just the ratio of the predicted colors when the IMF is allowed to vary to when it is held fixed.  Since optical colors are not sensitive to IMF differences (e.g. Appendix~B in \citealt{Bernardi2022}), $\alpha_r/\alpha_g=1$. 

Since this $\alpha$ is what affects $R_{e,*}/R_e$, we expect the size effects in $g$- to be similar to those in $r$-, especially if the light profiles have similar shapes.  In practice, the light profiles are slightly different:  the half-light radius is known to be smaller in $r$- than in $g$- \cite[e.g.,][]{Bernardi2003}.  However, if all is self-consistent, then the $R_{e,*}-M_*$ relations should be identical, whether they were determined from the $r$-band photometry or $g$-.  We return to this point in the next section.

\section{Comparison with quiescent galaxies at high redshift}\label{sec:hiz}
We now consider how the $z\sim 0$ MaNGA size-mass correlations compare with those estimated at $z\sim 1$ and $z\sim 2$.  We begin with the $R_e-M_*$ relations of quiescent galaxies in CANDELS over a range of $z$ \cite[from][]{Mowla2019}, because the $R_{e,*}-M_*$ relations of this same sample, with which we would like to compare, have been provided by \cite{Suess2019}.

\begin{figure}
  \centering
  \includegraphics[width=0.9\linewidth]{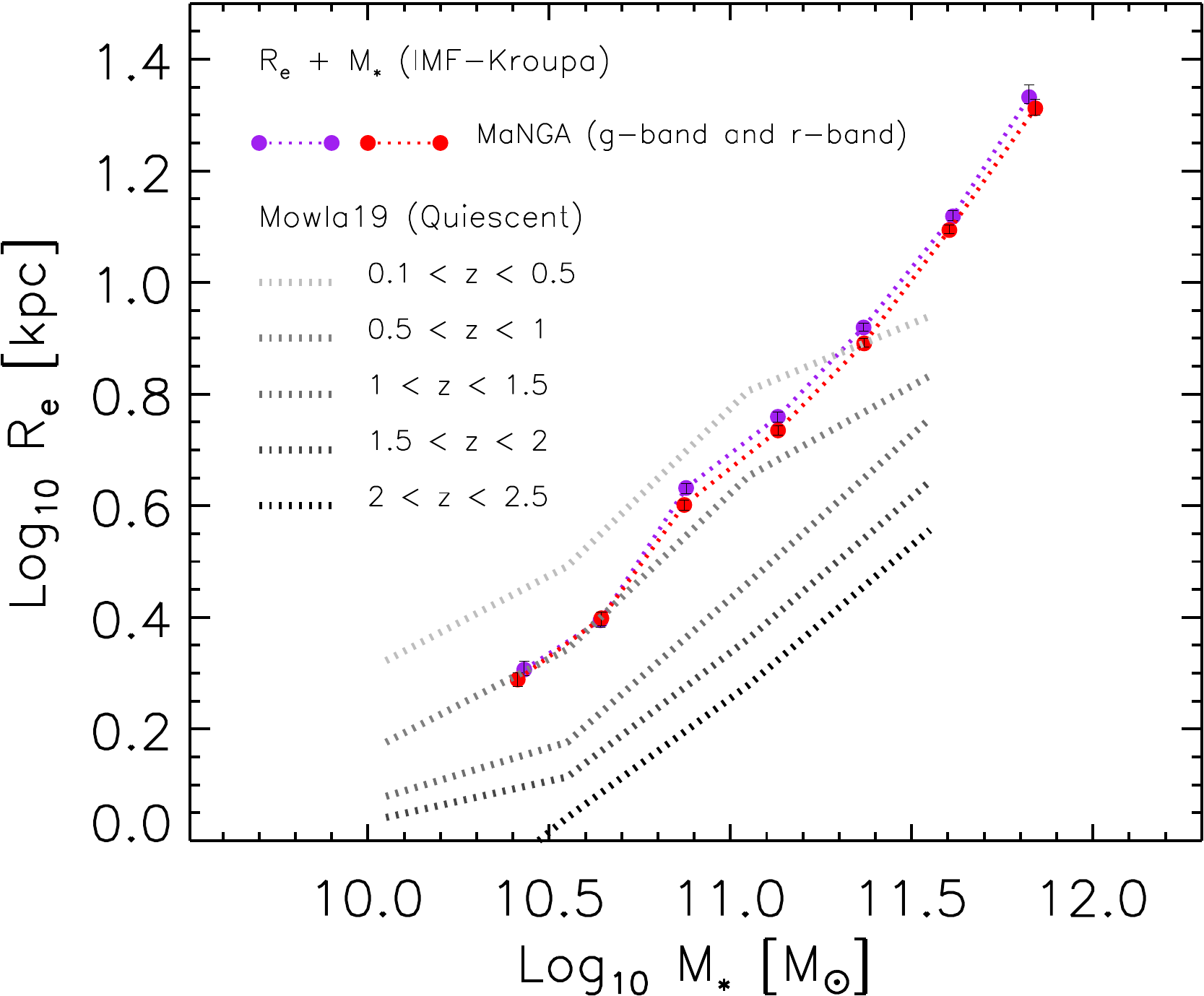}  
  \caption{Comparison of our MaNGA ($z\sim 0$) $R_e-M_*$ relation with corresponding relations at a range of other redshifts (from Mowla et al. 2019, as labeled).}
  \label{fig:mowla}
\end{figure}

The grey dotted lines in Figure~\ref{fig:mowla} show the $R_e-M_*$ relations of \cite{Mowla2019} in bins of width $\Delta z=0.5$ from $z=2.5$ to the present. (The high-$z$ sizes were corrected to restframe 5000~\AA, which lies between the SDSS $g$- and $r$-bands, and we have scaled the $M_*$ values from Chabrier- to Kroupa-IMF using the values provided in Table~2 of \citealt{Bernardi10}.)  Notice that the $R_e-M_*$ relation shows significant evolution from small sizes at $z\sim2$ to sizes that are $\sim 5\times$ larger at $z\sim 0$, in agreement with previous work (c.f. Introduction).
The red and purple symbols show our estimates of the $R_e-M_*$ relation in MaNGA (i.e. $z\sim 0$) in the $g$- and $r$-bands, which we showed are consistent with previous $z=0$ analyses.  The half-light radius in the $g$-band tends to be slightly larger (statisical errors are comparable to the symbol sizes); we return to this small difference shortly.  
The figure shows that the $z<0.5$ objects in \cite{Mowla2019} have {\em larger} half-light radii than MaNGA, especially at lower masses, suggesting that either there are systematic differences in the size estimates of the two samples, or that quiescent CANDELS galaxies are a rather different set of galaxies from MaNGA early-types. In principle, redshift-dependent selection effects, systematic uncertainties and/or progenitor bias could have biased the CANDELS $R_e$ evolution estimates \cite[e.g.][]{vanDokkumFranx1996, vanderWel2009, Shankar2015, Zanisi2021}.  However, Figure~\ref{fig:mowla} suggests that these issues may be less of a concern at high masses.

We now turn to the corresponding $R_{e,*}-M_*$ relations.  
The black curves in Figure~\ref{fig:hiz} show $R_{e,*}-M_*$ at $z\sim 1$ and 2 from \cite{Suess2019}.  They start from the grey curves in Figure~\ref{fig:mowla}, estimate $M_*/L$ gradients from optical color gradients at those redshifts, and use these to transform the $R_e$ values into $R_{e,*}$ values. While the higher redshift relations do allow for some $M_*/L$ gradients, they do not -- in fact, because they only use color information, they {\em cannot} -- account for IMF-driven gradients (or, for that matter, break degeneracies between age, metallicity and dust).  At each $z$, the $R_{e,*}-M_*$ relation is offset to smaller sizes than $R_e-M_*$.
How much the $R_{e,*}-M_*$ relation evolves depends on what one believes it is at $z\sim 0$.

The purple and red dotted curves in Figure~\ref{fig:hiz} show the $g$- and $r$-band $R_e-M_*$ relations of our MaNGA sample (same as previous figure), and the two solid curves show the corresponding $R_{e,*}-M_*$ relations.  Notice that the small dependence on waveband is reduced when going from $R_{e}-M_*$ to $R_{e,*}-M_*$.  
  This is a non-trivial and reassuring check on the self-consistency of our analysis, because the $M_*/L$ gradients are different in the two bands, as are the surface brightness profiles themselves.  
 \cite{Bernardi2022} discuss a number of systematic effects which can bias the $M_*/L$ values (see also right hand panel of Figure~\ref{fig:MsRe} and related discussion), but the important point is that the overall $R_{e,*}-M_*$ relation is offset to smaller sizes by significantly more than such systematics. 

\subsection{Implications for evolution}\label{sec:imply}
  When compared to the higher $z$ estimates of \cite{Suess2019}, our calibration in MaNGA, which accounts for IMF-driven gradients, suggests that $R_{e,*}-M_*$ evolves more from from $z\sim 2$ to $z\sim 1$ ($\sim 2$~Gyrs) than it does from $z\sim 1$ to $z\sim 0$ ($\sim 8$~Gyrs) though this depends slightly on mass.  On the other hand, if we fix the IMF in MaNGA (e.g. to Kroupa), then the $R_{e,*}$ values are about 0.2~dex higher (compare dashed and solid curves in left hand panel of Figure~\ref{fig:MsRe}), so we would conclude $R_{e,*}$ values have increased significantly since $z\sim 1$.

\begin{figure}
  \centering
  \includegraphics[width=0.9\linewidth]{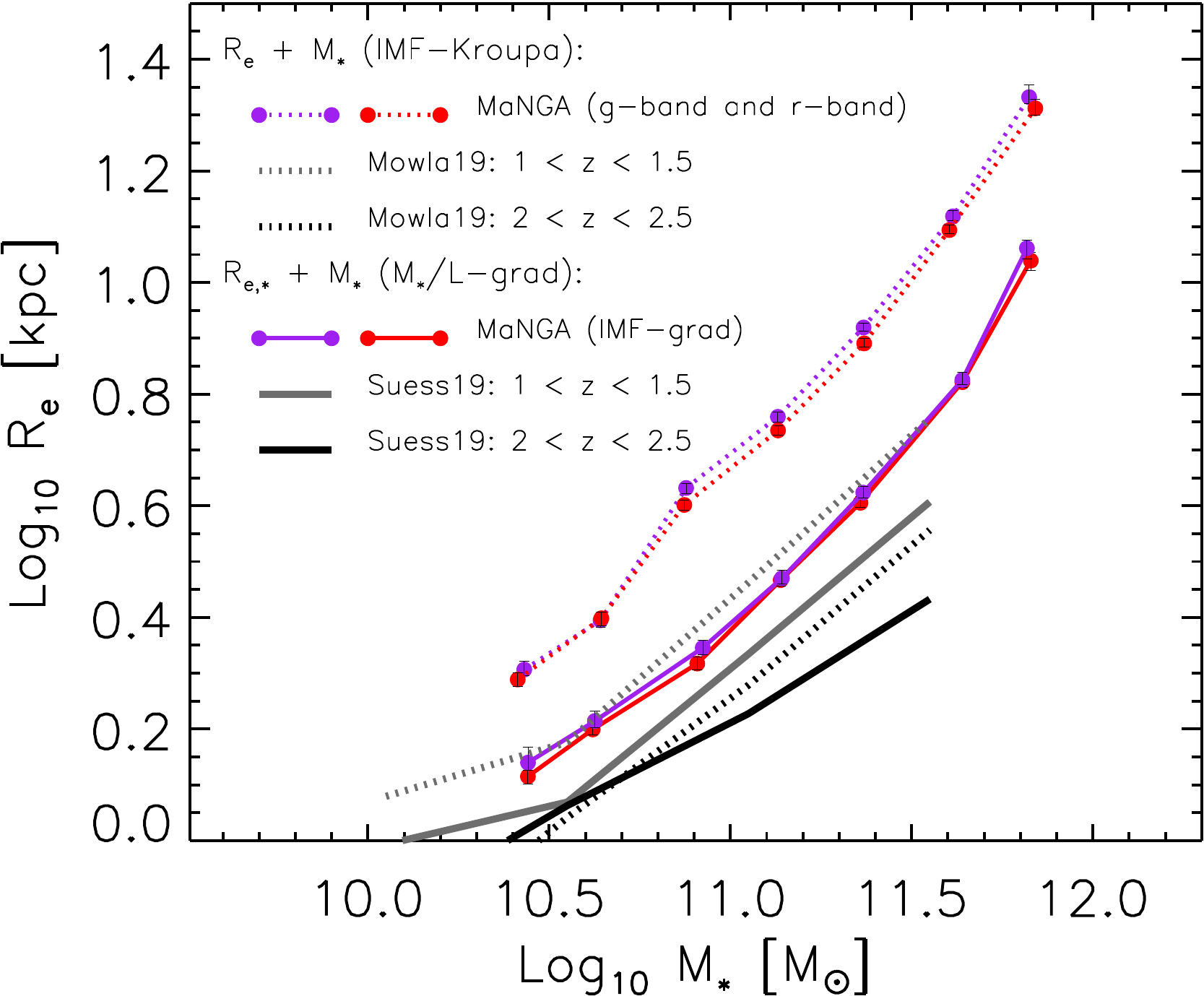}  
  \caption{Comparison of our $z\sim 0$ $R_e-M_*$ (fixed IMF) and $R_{e,*}-M_*$ (IMF-gradients) relations with recent estimates of the size-mass correlation at higher $z$ (from Mowla et al. 2019 and Suess et al. 2019a, with stellar masses scaled to Kroupa). Although the higher redshift relations allow for some $M_*/L$ gradients, they do not account for IMF-driven gradients.}
  \label{fig:hiz}
\end{figure}

We can express the evolution in terms of the ratio $R_{e,*}/R_e$ (bearing in mind that $R_e$ evolves significantly).  Comparing the high-$z$ solid and dotted curves in Figure~\ref{fig:hiz} shows that, at $M_*\sim 10^{11}M_\odot$, this ratio is closer to unity at $z\sim 2$ than it is at $z\sim 1$ ($R_{e,*}/R_e \sim 0.7$; see Fig. 6 in \citealt{Suess2019}).  If our IMF-gradient calibration is correct, it decreases even further as $z\to 0$ ($R_{e,*}/R_e \sim 0.6$).  However, if there are no IMF-driven gradients, then there must be a reversal in the trend towards $z\sim 0$, so that $R_{e,*}/R_e\sim 0.9$ or larger at $z\sim 0$ (see bottom panel of Figure~\ref{fig:mangaA}, in agreement with, e.g., \citealt{Chan2016}; see also Fig. 6 in \citealt{Ibarra-Medel2022}).

To have weak evolution of $R_{e,*}$ despite the strong observed evolution of $R_{e}$, $R_{e,*}/R_e$ must decrease at low $z$.
Since a negative $M_*/L$ gradient (higher $M_*/L$ in the center) will make $R_{e,*}$ smaller than $R_e$, a decrease of $R_{e,*}/R_e$ at low $z$ requires steeper (more negative) $M_*/L$ gradients at $z\sim 0$.  Our work has shown that there are two distinct things which can make this easier to achieve.  The first has to do with the stellar population:  at $z=0$, gradients in IMF lead to stronger gradients in $M_*/L$ (Figures~\ref{fig:profileE} and \ref{fig:mangaA}).
The second has to do with structure: the same $M_*/L$ gradient results in a smaller $R_{e,*}/R_e$ if $n$ is larger (Figures~\ref{fig:ratio-n} and \ref{fig:mangaBTn}).

So we are interested in how gradients and structure evolve, and the effect this has on $R_{e,*}/R_e$.  In this context, it is interesting to first consider a passively evolving population with an age gradient.  The age gradient will result in an evolving $M_*/L$ gradient, so $R_{e,*}/R_e$ will also evolve.  If quenching is `inside-out', so the center is older, then the outer parts fade more rapidly, so $R_e$ will decrease and $R_{e,*}/R_e$ will increase as the galaxy ages (passive evolution means $M_*$ and $R_{e,*}$ are fixed).  \cite{Ibarra-Medel2022} used this type of model to explain the increase of $R_{e,*}/R_e$ from the $z=0.5$ value ($R_{e,*}/R_e\sim 0.7$) to match what they believe is the $z=0$ value ($R_{e,*}/R_e\sim 0.9$, a value we would claim is biased high because it ignores IMF-gradients at $z=0$).  However, this has $R_e$ at $z>0$ {\em larger} than at $z=0$, which observations (e.g. Figure~\ref{fig:mowla}) have ruled out. (Indeed, as \citealt{LaBarbera2009} noted, for passive evolution models to reproduce the decrease of $R_e$ with increasing $z$, galaxies must be younger in their central regions.)  Therefore, to have {\em both} $R_e$ and $R_{e,*}/R_e$ increase between high-$z$ and $z=0$, there must have been a series of minor mergers of objects with large $M_*/L$, so that the mass added to the outskirts increases $R_{e,*}$ more than $R_e$, and offsets the passive evolution trend to have $R_e$ decrease.  However, such mergers will tend to increase $n$ \cite[e.g.][]{vD2010, Hilz2013, Shankar2018}.  If $n$ is larger, and the overall $M_*/L$ gradient is still negative, then the same negative $M_*/L$ gradient has a larger effect on $R_{e,*}/R_e$ (makes it smaller).  So, it appears that having both $R_e$ and $R_{e,*}/R_e$ increase as $z\to 0$ is difficult; the smaller $R_{e,*}/R_e\sim 0.6$ values which our IMF-gradient driven analysis returns at $z\sim 0$ are more natural than estimates based on analyses which ignore IMF-gradients \cite[e.g.][]{Szomoru2013, Chan2016}.  

We turn now to the values of $M_*/L$ and $R_{e,*}/R_e$ at higher $z$.  The \cite{Suess2019} estimate of $R_{e,*}/R_e \sim 0.75$ at $z\sim 1.4$ is significantly larger than that of \cite{Chan2016} ($R_{e,*}/R_e \sim 0.6$), which is based on a smaller sample of cluster galaxies at the same $z$.  Even so, both these values are smaller than the $z=0$ value one infers if IMF-gradients are ignored ($R_{e,*}/R_e \sim 0.9$).  This requires steeper negative $M_*/L$ gradients at high-$z$ than $z\sim 0$, and would imply a similar or even more dramatic evolution in half-mass radius $R_{e,*}$ than is observed for $R_e$ (between $z\sim 1.4$ and 0).  For the reasons given in the previous paragraph, minor mergers will have a more difficult time to produce $R_{e,*}/R_e\sim 0.9$ than our IMF-gradient based value $R_{e,*}/R_e\sim 0.6$ at $z=0$.

If IMF-gradients at $z\sim 0$ are required to produce sensible $R_{e,*}$ estimates, it is reasonable to ask if the higher $z$ estimates, which ignore IMF-driven gradients, are biased.  Assuming that dust has not compromised their analyses, we can think of two extreme scenarios.  (i)  IMF-gradients were {\em not} present at high-$z$, but the IMF varied strongly across the high-$z$ population, with little variation within a galaxy.  Then, the IMF-gradients we see in MaNGA today must be the result of mergers:  e.g., cores having bottom-heavy IMFs accreted objects having more Kroupa-like IMFs in their outer parts.  However, major mergers usually do not create gradients, and minor mergers are usually thought to only affect the outer regions.  Since the gradient we see at $z\sim 0$ is confined to scales {\em smaller} than $R_e$, minor mergers must have affected these smaller scales as well.  
If these same mergers increase $n$, then both effects act together to decrease $R_{e,*}/R_e$ to the $\sim 0.6$ value we estimate at $z\sim 0$.
(ii) Alternatively, if IMF-gradients were already in place in massive galaxies at e.g., $z\sim 2$, then we must estimate the additional effect they have on $M_*/L$ gradients if age gradients are also present. 
An age change of 1~Gyr produces a larger fractional change in $M_*/L$ when the galaxy is 3~Gyrs old than when it is 9 \citep{Tinsley1972}. In contrast, the fractional change in $M_*/L$ between Kroupa and Salpeter IMFs is approximately independent of age.  So, for $M_*/L$, the same age difference matters more when the galaxy is young.  
This raises the question of the sign of the age gradient at high-$z$.  In MaNGA (at $z \sim 0$), we find the central regions are slightly {\em younger} and more metal rich \cite[in agreement with SAMI;][]{Santucci2020}.  Hence, if a positive age gradient was present in massive galaxies at high-$z$, a passive evolution model similar to that of \cite{LaBarbera2009}, but with age and IMF gradients contributing with opposite signs at higher $z$ and color gradients dominated by metallicity effects, could explain the observed evolution: $R_e$ would increase passively as the object ages (i.e., without mergers). Minor mergers will provide additional increase. Thus, at $z>1$, gradients in optical color may fortuitously have captured the dominant $M_*/L$ trends, so the \cite{Suess2019} or \cite{Chan2016} estimates may not be too biased. If, instead, the central parts are older (e.g., \citealt{Chan2016}), then the age and IMF gradients add to steepen the $M_*/L$ gradient (i.e. more negative), so $R_{*,e}/R_e$ is even smaller than \cite{Suess2019} or \cite{Chan2016} assume.  In this case, more merging is {\em required} to increase $R_{e,*}$. In any case, the required merging between high-$z$ and the present is less extreme if $R_{e,*}/R_e$ at $z\sim0$ is closer to our calibration ($R_{e,*}/R_e \sim 0.6$) than to unity.

In conclusion:  While it is tempting to conclude that even though there has been been substantial evolution in the $R_e-M_*$ relation of early-type/quiescent galaxies there may have been little evolution in the $R_{e,*}-M_*$ relation, this conclusion is only correct if $M_*/L$ gradients at high-$z$, whether age- and/or IMF-driven, are weaker than at $z\sim 0$ and/or the light profile steepens at low $z$.  While these are all plausible, and there is some evidence of the latter \cite[e.g.][]{vD2010, Shankar2018}, the former has not yet been consistently proven.  This obviously matters greatly for studies which seek to use the evolution of galaxy sizes to constrain assembly histories \cite[e.g.][]{Hopkins2010, Shankar2013, Shankar2015, Zanisi2021}.

\section{Conclusions}
We studied the dependence of the size-mass relation of early-type galaxies in the MaNGA survey on how the sizes and masses were estimated.  Specifically, we addressed the question of how the projected half-mass radius differs from the half-light radius.  The $M_*/L$ gradient which might cause this can arise from age gradients even if the IMF is constant throughout a galaxy, or from IMF gradients, or both (equation~\ref{eq:R*} and related discussion).  As age gradients in old quiescent galaxies at $z\sim 0$ are small, IMF gradients may be the dominant cause of $M_*/L$ gradients in low-$z$ early-type galaxies.  

The effect of such gradients on the size estimate depends on the surface brightness profile: when expressed in terms of $R_e$ (e.g. equation~\ref{eq:toy}), the same $M_*/L$ gradient has a much larger effect if the S{\'e}rsic index $n$ is large (Figure~\ref{fig:ratio-n}).  However, the importance of $n$ can be masked by scatter in the $M_*/L$ gradients.  Gradients which extend to a smaller fraction of $R_e$ have a smaller effect (Figure~\ref{fig:compareLB19-1}). 

We used the IMF shapes determined self-consistently along with other simple stellar population parameters (age, metallicity and $\alpha$-enhancement) from stacked spectra of early-type galaxies in the MaNGA survey by \cite{Bernardi2022}. The objects in the sample span a wide range of stellar masses and surface brightness profiles (Figures~\ref{fig:flagM*} and~\ref{fig:morphM*}), but the sample is sufficiently large that they can be stacked in relatively narrow bins in luminosity, velocity dispersion and morphological type (Table~\ref{tab:Cat}).

The IMF tends to be more bottom heavy in the central regions compared to beyond $R_e$ (Figure~\ref{fig:profileE}). The corresponding $M_*/L$ gradients are significantly stronger than those obtained at fixed IMF (e.g. Kroupa).  The IMF-driven $M_*/L$ gradients in MaNGA early-type galaxies (Figure~\ref{fig:profileE} and Table~\ref{tab:Cat}) tend to slightly increase the inferred stellar mass estimate but decrease the projected half-mass radius more substantially, by an amount that depends on S{\'e}rsic index $n$ (Figures~\ref{fig:mangaBTn}--\ref{fig:mangaA}).  

The $R_{e,*}-M_*$ relation which results from accounting for IMF-driven gradients is shifted towards smaller sizes by almost 0.3~dex compared to the $R_e-M_*$ relation estimated from the half-light radius and the fixed (Kroupa) IMF estimate of $M_*$ (Figure~\ref{fig:MsRe} and Table~\ref{tab:RMfits}). One gets similar results whether starting from photometry in the $g$- or the $r$-band:  The $R_{e,*}/R_e$ ratio is similar in the two bands (Figure~\ref{fig:alphagr}) as is the $R_{e,*}-M_*$ relation which results (Figure~\ref{fig:hiz}).

In MaNGA, we only see significant differences between $R_e-M_*$ and $R_{e,*}-M_*$ if we include IMF-driven gradients (Figure~\ref{fig:MsRe}).  This complicates comparison with the $R_{e,*}-M_*$ relation at higher redshifts, since these higher-$z$ estimates either ignore, or are based on methods which are insensitive to, IMF-related effects (Figure~\ref{fig:hiz}). On the other hand, we noted that, in the younger stellar populations at higher $z$, age gradients may matter more when computing $M_*/L$ than at $z\sim 0$ (Section~\ref{sec:imply}).

While it is tempting to conclude that there may have been little evolution in the $R_{e,*}-M_*$ relation, this conclusion is only correct if $M_*/L$ gradients, whether age- and/or IMF-driven, at high-$z$ are weaker than at $z\sim 0$ and/or the light profile steepens at low $z$. Our results show that whatever the case at high-$z$, the required merging between high-$z$ and the present is less extreme if $R_{e,*}/R_e$ at $z\sim 0$ is closer to our calibration ($\sim 0.6$) than to unity (Section~\ref{sec:imply}).

We have concentrated on {\em gradients} in the IMF, and their impact on $R_{e,*}/R_e$ and its evolution.  While there is no observational consensus on the shape of the high-$z$ IMF \cite[see][for recent progress]{Mendel2020}, models which relate the shape of the IMF to the star formation rate \cite[e.g][]{Durham2016, Fontanot2020} predict that the typical IMF evolves.  It is a shallow power-law at the large SFRs that are more typical at high $z$; at lower SFRs (hence lower $z$), it bends from this power law towards smaller abundances at the largest and smallest masses.  Other recent work, which connects the IMF of low and intermediate mass stars to metallicity, and that of higher mass stars to both metallicity and environment, also concludes that the typical IMF evolves \citep{Jevrabkova2018, Yan2021, Sharda2022}.  We hope that our work inspires a study of the $M_*/L$ {\em gradients} in these models.  Likewise, in the EAGLE simulations of \cite{Barber2019}, the IMF depends on pressure, and hence on the SFR surface density, and so the typical IMF is predicted to evolve, giving rise to IMF gradients.  In their models, for older stellar populations, much of the resulting $M_*/L$ gradient is driven by these IMF, rather than age, gradients.  When coupled with a study of how the structural parameters of the objects evolves, this will provide an estimate of the expected impact on $R_{e,*}/R_e$ and its evolution. 

We end with a reminder that estimating the IMF is difficult even at low $z$.  A number of potential systematic effects, and the reasons for our particular analysis choices, are discussed in \cite{Bernardi2022}.  Nevertheless, because IMF gradients potentially impact conclusions about galaxy formation and assembly in non-trivial ways, we believe it is important that they be quantified.  Therefore, we are in the process of extending our analysis to include spirals.

\section*{Acknowledgements}
We are grateful to
K. Westfall for clarifications about changes in the MaNGA database between DR15 and DR17,
C. Conroy for discussion of SSP models, and
F. Shankar and the anonymous referee for detailed comments on the manuscript.  
This work was supported in part by NSF grant AST-1816330.

Funding for the Sloan Digital Sky Survey IV has been provided by the Alfred P. Sloan Foundation, the U.S. Department of Energy Office of Science, and the Participating Institutions. SDSS acknowledges support and resources from the Center for High-Performance Computing at the University of Utah. The SDSS web site is www.sdss.org.

SDSS is managed by the Astrophysical Research Consortium for the Participating Institutions of the SDSS Collaboration including the Brazilian Participation Group, the Carnegie Institution for Science, Carnegie Mellon University, the Chilean Participation Group, the French Participation Group, Harvard-Smithsonian Center for Astrophysics, Instituto de Astrof{\'i}sica de Canarias, The Johns Hopkins University, Kavli Institute for the Physics and Mathematics of the Universe (IPMU) / University of Tokyo, Lawrence Berkeley National Laboratory, Leibniz Institut f{\"u}r Astrophysik Potsdam (AIP), Max-Planck-Institut f{\"u}r Astronomie (MPIA Heidelberg), Max-Planck-Institut f{\"u}r Astrophysik (MPA Garching), Max-Planck-Institut f{\"u}r Extraterrestrische Physik (MPE), National Astronomical Observatories of China, New Mexico State University, New York University, University of Notre Dame, Observat{\'o}rio Nacional / MCTI, The Ohio State University, Pennsylvania State University, Shanghai Astronomical Observatory, United Kingdom Participation Group, Universidad Nacional Aut{\'o}noma de M{\'e}xico, University of Arizona, University of Colorado Boulder, University of Oxford, University of Portsmouth, University of Utah, University of Virginia, University of Washington, University of Wisconsin, Vanderbilt University, and Yale University.

\section*{Data availability}

The data underlying this article are available in the Sloan Digital Sky Survey Database at https://www.sdss.org/dr17/.





\bibliographystyle{mnras}
\bibliography{biblio} 

\begin{thebibliography}{}
\makeatletter
\relax
\def\mn@urlcharsother{\let\do\@makeother \do\$\do\&\do\#\do\^\do\_\do\%\do\~}
\def\mn@doi{\begingroup\mn@urlcharsother \@ifnextchar [ {\mn@doi@}
  {\mn@doi@[]}}
\def\mn@doi@[#1]#2{\def\@tempa{#1}\ifx\@tempa\@empty \href
  {http://dx.doi.org/#2} {doi:#2}\else \href {http://dx.doi.org/#2} {#1}\fi
  \endgroup}
\def\mn@eprint#1#2{\mn@eprint@#1:#2::\@nil}
\def\mn@eprint@arXiv#1{\href {http://arxiv.org/abs/#1} {{\tt arXiv:#1}}}
\def\mn@eprint@dblp#1{\href {http://dblp.uni-trier.de/rec/bibtex/#1.xml}
  {dblp:#1}}
\def\mn@eprint@#1:#2:#3:#4\@nil{\def\@tempa {#1}\def\@tempb {#2}\def\@tempc
  {#3}\ifx \@tempc \@empty \let \@tempc \@tempb \let \@tempb \@tempa \fi \ifx
  \@tempb \@empty \def\@tempb {arXiv}\fi \@ifundefined
  {mn@eprint@\@tempb}{\@tempb:\@tempc}{\expandafter \expandafter \csname
  mn@eprint@\@tempb\endcsname \expandafter{\@tempc}}}

\bibitem[\protect\citeauthoryear{{Abdurro'uf} et~al.,}{{Abdurro'uf}
  et~al.}{2021}]{sdssDR17}
{Abdurro'uf} et~al., 2021, arXiv e-prints, \href
  {https://ui.adsabs.harvard.edu/abs/2021arXiv211202026A} {p. arXiv:2112.02026}

\bibitem[\protect\citeauthoryear{{Aguado} et~al.,}{{Aguado}
  et~al.}{2019}]{Aguado2019}
{Aguado} D.~S.,  et~al., 2019, \mn@doi [\apjs] {10.3847/1538-4365/aaf651},
  \href {http://adsabs.harvard.edu/abs/2019ApJS..240...23A} {240, 23}

\bibitem[\protect\citeauthoryear{{Barber}, {Schaye}  \& {Crain}}{{Barber}
  et~al.}{2019}]{Barber2019}
{Barber} C.,  {Schaye} J.,   {Crain} R.~A.,  2019, \mn@doi [\mnras]
  {10.1093/mnras/sty3011}, \href
  {https://ui.adsabs.harvard.edu/abs/2019MNRAS.483..985B} {483, 985}

\bibitem[\protect\citeauthoryear{{Barro} et~al.,}{{Barro}
  et~al.}{2017}]{Barro2017}
{Barro} G.,  et~al., 2017, \mn@doi [\apj] {10.3847/1538-4357/aa6b05}, \href
  {https://ui.adsabs.harvard.edu/abs/2017ApJ...840...47B} {840, 47}

\bibitem[\protect\citeauthoryear{{Bernardi} et~al.,}{{Bernardi}
  et~al.}{2003}]{Bernardi2003}
{Bernardi} M.,  et~al., 2003, \mn@doi [\aj] {10.1086/374256}, \href
  {https://ui.adsabs.harvard.edu/abs/2003AJ....125.1849B} {125, 1849}

\bibitem[\protect\citeauthoryear{{Bernardi}, {Shankar}, {Hyde}, {Mei},
  {Marulli}  \& {Sheth}}{{Bernardi} et~al.}{2010}]{Bernardi10}
{Bernardi} M.,  {Shankar} F.,  {Hyde} J.~B.,  {Mei} S.,  {Marulli} F.,
  {Sheth} R.~K.,  2010, \mn@doi [\mnras] {10.1111/j.1365-2966.2010.16425.x},
  \href {http://adsabs.harvard.edu/abs/2010MNRAS.404.2087B} {404, 2087}

\bibitem[\protect\citeauthoryear{{Bernardi}, {Meert}, {Vikram},
  {Huertas-Company}, {Mei}, {Shankar}  \& {Sheth}}{{Bernardi}
  et~al.}{2014}]{Bernardi2014}
{Bernardi} M.,  {Meert} A.,  {Vikram} V.,  {Huertas-Company} M.,  {Mei} S.,
  {Shankar} F.,   {Sheth} R.~K.,  2014, \mn@doi [\mnras]
  {10.1093/mnras/stu1106}, \href
  {http://adsabs.harvard.edu/abs/2014MNRAS.443..874B} {443, 874}

\bibitem[\protect\citeauthoryear{{Bernardi}, {Sheth}, {Dominguez-Sanchez},
  {Fischer}, {Chae}, {Huertas-Company}  \& {Shankar}}{{Bernardi}
  et~al.}{2018}]{Bernardi2018b}
{Bernardi} M.,  {Sheth} R.~K.,  {Dominguez-Sanchez} H.,  {Fischer} J.-L.,
  {Chae} K.-H.,  {Huertas-Company} M.,   {Shankar} F.,  2018, \mn@doi [\mnras]
  {10.1093/mnras/sty781}, \href
  {http://adsabs.harvard.edu/abs/2018MNRAS.477.2560B} {477, 2560}

\bibitem[\protect\citeauthoryear{{Bernardi}, {Dom{\'\i}nguez S{\'a}nchez},
  {Brownstein}, {Drory}  \& {Sheth}}{{Bernardi} et~al.}{2019}]{Bernardi2019}
{Bernardi} M.,  {Dom{\'\i}nguez S{\'a}nchez} H.,  {Brownstein} J.~R.,  {Drory}
  N.,   {Sheth} R.~K.,  2019, \mn@doi [\mnras] {10.1093/mnras/stz2413}, \href
  {https://ui.adsabs.harvard.edu/abs/2019MNRAS.489.5633B} {489, 5633}

\bibitem[\protect\citeauthoryear{{Bernardi}, {Dom{\'\i}nguez S{\'a}nchez},
  {Margalef-Bentabol}, {Nikakhtar}  \& {Sheth}}{{Bernardi}
  et~al.}{2020}]{Bernardi2020}
{Bernardi} M.,  {Dom{\'\i}nguez S{\'a}nchez} H.,  {Margalef-Bentabol} B.,
  {Nikakhtar} F.,   {Sheth} R.~K.,  2020, \mn@doi [\mnras]
  {10.1093/mnras/staa1064}, \href
  {https://ui.adsabs.harvard.edu/abs/2020MNRAS.494.5148B} {494, 5148}

\bibitem[\protect\citeauthoryear{{Bernardi}, {Dom{\'i}nguez S{\'a}nchez},
  {Sheth}, {Brownstein}  \& {Lane}}{{Bernardi} et~al.}{2022}]{Bernardi2022}
{Bernardi} M.,  {Dom{\'i}nguez S{\'a}nchez} H.,  {Sheth} R.~K.,  {Brownstein}
  J.~R.,   {Lane} R.~R.,  2022, arXiv e-prints, \href
  {https://ui.adsabs.harvard.edu/abs/2022arXiv221105800B} {p. arXiv:2211.05800}

\bibitem[\protect\citeauthoryear{{Blanton} et~al.,}{{Blanton}
  et~al.}{2017}]{Blanton2017}
{Blanton} M.~R.,  et~al., 2017, \mn@doi [\aj] {10.3847/1538-3881/aa7567}, \href
  {http://adsabs.harvard.edu/abs/2017AJ....154...28B} {154, 28}

\bibitem[\protect\citeauthoryear{{Buitrago}, {Trujillo}, {Conselice},
  {Bouwens}, {Dickinson}  \& {Yan}}{{Buitrago} et~al.}{2008}]{Buitrago2008}
{Buitrago} F.,  {Trujillo} I.,  {Conselice} C.~J.,  {Bouwens} R.~J.,
  {Dickinson} M.,   {Yan} H.,  2008, \mn@doi [\apjl] {10.1086/592836}, \href
  {https://ui.adsabs.harvard.edu/abs/2008ApJ...687L..61B} {687, L61}

\bibitem[\protect\citeauthoryear{{Bundy} et~al.,}{{Bundy}
  et~al.}{2015}]{Bundy2015}
{Bundy} K.,  et~al., 2015, \mn@doi [\apj] {10.1088/0004-637X/798/1/7}, \href
  {http://adsabs.harvard.edu/abs/2015ApJ...798....7B} {798, 7}

\bibitem[\protect\citeauthoryear{{Cappellari} et~al.,}{{Cappellari}
  et~al.}{2013}]{Cappellari2013a}
{Cappellari} M.,  et~al., 2013, \mn@doi [\mnras] {10.1093/mnras/stt562}, \href
  {http://adsabs.harvard.edu/abs/2013MNRAS.432.1709C} {432, 1709}

\bibitem[\protect\citeauthoryear{{Chan} et~al.,}{{Chan}
  et~al.}{2016}]{Chan2016}
{Chan} J. C.~C.,  et~al., 2016, \mn@doi [\mnras] {10.1093/mnras/stw502}, \href
  {https://ui.adsabs.harvard.edu/abs/2016MNRAS.458.3181C} {458, 3181}

\bibitem[\protect\citeauthoryear{{Cimatti} et~al.,}{{Cimatti}
  et~al.}{2008}]{Cimatti2008}
{Cimatti} A.,  et~al., 2008, \mn@doi [\aap] {10.1051/0004-6361:20078739}, \href
  {https://ui.adsabs.harvard.edu/abs/2008A&A...482...21C} {482, 21}

\bibitem[\protect\citeauthoryear{{Daddi} et~al.,}{{Daddi}
  et~al.}{2005}]{Daddi2005}
{Daddi} E.,  et~al., 2005, \mn@doi [\apj] {10.1086/430104}, \href
  {https://ui.adsabs.harvard.edu/abs/2005ApJ...626..680D} {626, 680}

\bibitem[\protect\citeauthoryear{{Dom{\'\i}nguez S{\'a}nchez}, {Bernardi},
  {Brownstein}, {Drory}  \& {Sheth}}{{Dom{\'\i}nguez S{\'a}nchez}
  et~al.}{2019}]{DS2019}
{Dom{\'\i}nguez S{\'a}nchez} H.,  {Bernardi} M.,  {Brownstein} J.~R.,  {Drory}
  N.,   {Sheth} R.~K.,  2019, \mn@doi [\mnras] {10.1093/mnras/stz2414}, \href
  {https://ui.adsabs.harvard.edu/abs/2019MNRAS.489.5612D} {489, 5612}

\bibitem[\protect\citeauthoryear{{Dom{\'\i}nguez S{\'a}nchez}, {Bernardi},
  {Nikakhtar}, {Margalef-Bentabol}  \& {Sheth}}{{Dom{\'\i}nguez S{\'a}nchez}
  et~al.}{2020}]{DS2020}
{Dom{\'\i}nguez S{\'a}nchez} H.,  {Bernardi} M.,  {Nikakhtar} F.,
  {Margalef-Bentabol} B.,   {Sheth} R.~K.,  2020, \mn@doi [\mnras]
  {10.1093/mnras/staa1364}, \href
  {https://ui.adsabs.harvard.edu/abs/2020MNRAS.495.2894D} {495, 2894}

\bibitem[\protect\citeauthoryear{{Dom{\'\i}nguez S{\'a}nchez}, {Margalef},
  {Bernardi}  \& {Huertas-Company}}{{Dom{\'\i}nguez S{\'a}nchez}
  et~al.}{2021}]{DS21}
{Dom{\'\i}nguez S{\'a}nchez} H.,  {Margalef} B.,  {Bernardi} M.,
  {Huertas-Company} M.,  2021, \mn@doi [\mnras] {10.1093/mnras/stab3089}, \href
  {https://ui.adsabs.harvard.edu/abs/2021MNRAS.tmp.2822D} {}

\bibitem[\protect\citeauthoryear{{Drory} et~al.,}{{Drory}
  et~al.}{2015}]{Drory2015}
{Drory} N.,  et~al., 2015, \mn@doi [\aj] {10.1088/0004-6256/149/2/77}, \href
  {http://adsabs.harvard.edu/abs/2015AJ....149...77D} {149, 77}

\bibitem[\protect\citeauthoryear{{Emsellem} et~al.,}{{Emsellem}
  et~al.}{2007}]{Emsellem2007}
{Emsellem} E.,  et~al., 2007, \mn@doi [\mnras]
  {10.1111/j.1365-2966.2007.11752.x}, \href
  {http://adsabs.harvard.edu/abs/2007MNRAS.379..401E} {379, 401}

\bibitem[\protect\citeauthoryear{{Feldmeier-Krause}, {Lonoce}  \&
  {Freedman}}{{Feldmeier-Krause} et~al.}{2021}]{FeldmeierKrause2021}
{Feldmeier-Krause} A.,  {Lonoce} I.,   {Freedman} W.~L.,  2021, arXiv e-prints,
  \href {https://ui.adsabs.harvard.edu/abs/2021arXiv211002860F} {p.
  arXiv:2110.02860}

\bibitem[\protect\citeauthoryear{{Fischer}, {Dom{\'{\i}}nguez S{\'a}nchez}  \&
  {Bernardi}}{{Fischer} et~al.}{2019}]{Fischer2019}
{Fischer} J.-L.,  {Dom{\'{\i}}nguez S{\'a}nchez} H.,   {Bernardi} M.,  2019,
  \mn@doi [\mnras] {10.1093/mnras/sty3135}, \href
  {http://adsabs.harvard.edu/abs/2019MNRAS.483.2057F} {483, 2057}

\bibitem[\protect\citeauthoryear{{Fontanot}}{{Fontanot}}{2020}]{Fontanot2020}
{Fontanot} F.,  2020, in {Boquien} M.,  {Lusso} E.,  {Gruppioni} C.,
  {Tissera} P.,  eds,  Vol. 341, Panchromatic Modelling with Next Generation
  Facilities. pp 124--128 (\mn@eprint {arXiv} {1903.03647}),
  \mn@doi{10.1017/S1743921319002370}

\bibitem[\protect\citeauthoryear{{Garc{\'\i}a-Benito}, {Gonz{\'a}lez Delgado},
  {P{\'e}rez}, {Cid Fernandes}, {S{\'a}nchez}  \& {de
  Amorim}}{{Garc{\'\i}a-Benito} et~al.}{2019}]{califa2019}
{Garc{\'\i}a-Benito} R.,  {Gonz{\'a}lez Delgado} R.~M.,  {P{\'e}rez} E.,  {Cid
  Fernandes} R.,  {S{\'a}nchez} S.~F.,   {de Amorim} A.~L.,  2019, \mn@doi
  [\aap] {10.1051/0004-6361/201833993}, \href
  {https://ui.adsabs.harvard.edu/abs/2019A&A...621A.120G} {621, A120}

\bibitem[\protect\citeauthoryear{{Ge}, {Mao}, {Lu}, {Cappellari}, {Long}  \&
  {Yan}}{{Ge} et~al.}{2021}]{Ge2021}
{Ge} J.,  {Mao} S.,  {Lu} Y.,  {Cappellari} M.,  {Long} R.~J.,   {Yan} R.,
  2021, \mn@doi [\mnras] {10.1093/mnras/stab2341}, \href
  {https://ui.adsabs.harvard.edu/abs/2021MNRAS.507.2488G} {507, 2488}

\bibitem[\protect\citeauthoryear{{Girardi}, {Bressan}, {Bertelli}  \&
  {Chiosi}}{{Girardi} et~al.}{2000}]{Girardi2000}
{Girardi} L.,  {Bressan} A.,  {Bertelli} G.,   {Chiosi} C.,  2000, \mn@doi
  [\aaps] {10.1051/aas:2000126}, \href
  {http://adsabs.harvard.edu/abs/2000A%26AS..141..371G} {141, 371}

\bibitem[\protect\citeauthoryear{{Graham} et~al.,}{{Graham}
  et~al.}{2018}]{Graham2018}
{Graham} M.~T.,  et~al., 2018, \mn@doi [\mnras] {10.1093/mnras/sty504}, \href
  {http://adsabs.harvard.edu/abs/2018MNRAS.477.4711G} {477, 4711}

\bibitem[\protect\citeauthoryear{{Gunn} et~al.,}{{Gunn}
  et~al.}{2006}]{Gunn2006}
{Gunn} J.~E.,  et~al., 2006, \mn@doi [\aj] {10.1086/500975}, \href
  {https://ui.adsabs.harvard.edu/abs/2006AJ....131.2332G} {131, 2332}

\bibitem[\protect\citeauthoryear{{Hilz}, {Naab}  \& {Ostriker}}{{Hilz}
  et~al.}{2013}]{Hilz2013}
{Hilz} M.,  {Naab} T.,   {Ostriker} J.~P.,  2013, \mn@doi [\mnras]
  {10.1093/mnras/sts501}, \href
  {http://adsabs.harvard.edu/abs/2013MNRAS.429.2924H} {429, 2924}

\bibitem[\protect\citeauthoryear{{Hirschmann}, {Naab}, {Ostriker}, {Forbes},
  {Duc}, {Dav{\'e}}, {Oser}  \& {Karabal}}{{Hirschmann}
  et~al.}{2015}]{Hirschmann2015}
{Hirschmann} M.,  {Naab} T.,  {Ostriker} J.~P.,  {Forbes} D.~A.,  {Duc} P.-A.,
  {Dav{\'e}} R.,  {Oser} L.,   {Karabal} E.,  2015, \mn@doi [\mnras]
  {10.1093/mnras/stv274}, \href
  {http://adsabs.harvard.edu/abs/2015MNRAS.449..528H} {449, 528}

\bibitem[\protect\citeauthoryear{{Hopkins}, {Bundy}, {Hernquist}, {Wuyts}  \&
  {Cox}}{{Hopkins} et~al.}{2010}]{Hopkins2010}
{Hopkins} P.~F.,  {Bundy} K.,  {Hernquist} L.,  {Wuyts} S.,   {Cox} T.~J.,
  2010, \mn@doi [\mnras] {10.1111/j.1365-2966.2009.15699.x}, \href
  {https://ui.adsabs.harvard.edu/abs/2010MNRAS.401.1099H} {401, 1099}

\bibitem[\protect\citeauthoryear{{Ibarra-Medel}, {Avila-Reese}, {Lacerna},
  {Rodr{\'\i}guez-Puebla}, {V{\'a}zquez-Mata}, {Hern{\'a}ndez-Toledo}  \&
  {S{\'a}nchez}}{{Ibarra-Medel} et~al.}{2021}]{Ibarra-Medel2022}
{Ibarra-Medel} H.,  {Avila-Reese} V.,  {Lacerna} I.,  {Rodr{\'\i}guez-Puebla}
  A.,  {V{\'a}zquez-Mata} J.~A.,  {Hern{\'a}ndez-Toledo} H.~M.,   {S{\'a}nchez}
  S.~F.,  2021, arXiv e-prints, \href
  {https://ui.adsabs.harvard.edu/abs/2021arXiv211212799I} {p. arXiv:2112.12799}

\bibitem[\protect\citeauthoryear{{Je{\v{r}}{\'a}bkov{\'a}}, {Hasani Zonoozi},
  {Kroupa}, {Beccari}, {Yan}, {Vazdekis}  \& {Zhang}}{{Je{\v{r}}{\'a}bkov{\'a}}
  et~al.}{2018}]{Jevrabkova2018}
{Je{\v{r}}{\'a}bkov{\'a}} T.,  {Hasani Zonoozi} A.,  {Kroupa} P.,  {Beccari}
  G.,  {Yan} Z.,  {Vazdekis} A.,   {Zhang} Z.~Y.,  2018, \mn@doi [\aap]
  {10.1051/0004-6361/201833055}, \href
  {https://ui.adsabs.harvard.edu/abs/2018A&A...620A..39J} {620, A39}

\bibitem[\protect\citeauthoryear{{Kennedy} et~al.,}{{Kennedy}
  et~al.}{2015}]{Kennedy2015}
{Kennedy} R.,  et~al., 2015, \mn@doi [\mnras] {10.1093/mnras/stv2032}, \href
  {https://ui.adsabs.harvard.edu/abs/2015MNRAS.454..806K} {454, 806}

\bibitem[\protect\citeauthoryear{{Kuntschner}}{{Kuntschner}}{2015}]{Kuntschner2015}
{Kuntschner} H.,  2015, in {Cappellari} M.,  {Courteau} S.,  eds,  Vol. 311,
  Galaxy Masses as Constraints of Formation Models. pp 53--56,
  \mn@doi{10.1017/S1743921315003385}

\bibitem[\protect\citeauthoryear{{La Barbera} \& {de Carvalho}}{{La Barbera} \&
  {de Carvalho}}{2009}]{LaBarbera2009}
{La Barbera} F.,  {de Carvalho} R.~R.,  2009, \mn@doi [\apjl]
  {10.1088/0004-637X/699/2/L76}, \href
  {https://ui.adsabs.harvard.edu/abs/2009ApJ...699L..76L} {699, L76}

\bibitem[\protect\citeauthoryear{{La Barbera}, {Ferreras}, {Vazdekis}, {de la
  Rosa}, {de Carvalho}, {Trevisan}, {Falc{\'o}n-Barroso}  \&
  {Ricciardelli}}{{La Barbera} et~al.}{2013}]{LaBarbera2013}
{La Barbera} F.,  {Ferreras} I.,  {Vazdekis} A.,  {de la Rosa} I.~G.,  {de
  Carvalho} R.~R.,  {Trevisan} M.,  {Falc{\'o}n-Barroso} J.,   {Ricciardelli}
  E.,  2013, \mn@doi [\mnras] {10.1093/mnras/stt943}, \href
  {http://adsabs.harvard.edu/abs/2013MNRAS.433.3017L} {433, 3017}

\bibitem[\protect\citeauthoryear{{La Barbera}, {Vazdekis}, {Ferreras},
  {Pasquali}, {Cappellari}, {Mart{\'{\i}}n-Navarro}, {Sch{\"o}nebeck}  \&
  {Falc{\'o}n-Barroso}}{{La Barbera} et~al.}{2016}]{LaBarbera2016}
{La Barbera} F.,  {Vazdekis} A.,  {Ferreras} I.,  {Pasquali} A.,  {Cappellari}
  M.,  {Mart{\'{\i}}n-Navarro} I.,  {Sch{\"o}nebeck} F.,   {Falc{\'o}n-Barroso}
  J.,  2016, \mn@doi [\mnras] {10.1093/mnras/stv2996}, \href
  {http://adsabs.harvard.edu/abs/2016MNRAS.457.1468L} {457, 1468}

\bibitem[\protect\citeauthoryear{{La Barbera}, {Vazdekis}, {Ferreras},
  {Pasquali}, {Allende Prieto}, {R{\"o}ck}, {Aguado}  \& {Peletier}}{{La
  Barbera} et~al.}{2017}]{LaBarbera2017}
{La Barbera} F.,  {Vazdekis} A.,  {Ferreras} I.,  {Pasquali} A.,  {Allende
  Prieto} C.,  {R{\"o}ck} B.,  {Aguado} D.~S.,   {Peletier} R.~F.,  2017,
  \mn@doi [\mnras] {10.1093/mnras/stw2407}, \href
  {https://ui.adsabs.harvard.edu/abs/2017MNRAS.464.3597L} {464, 3597}

\bibitem[\protect\citeauthoryear{{La Barbera} et~al.,}{{La Barbera}
  et~al.}{2019}]{LaBarbera2019}
{La Barbera} F.,  et~al., 2019, \mn@doi [\mnras] {10.1093/mnras/stz2192}, \href
  {https://ui.adsabs.harvard.edu/abs/2019MNRAS.489.4090L} {489, 4090}

\bibitem[\protect\citeauthoryear{{Lacerna}, {Ibarra-Medel}, {Avila-Reese},
  {Hern{\'a}ndez-Toledo}, {V{\'a}zquez-Mata}  \& {S{\'a}nchez}}{{Lacerna}
  et~al.}{2020}]{Lacerna2020}
{Lacerna} I.,  {Ibarra-Medel} H.,  {Avila-Reese} V.,  {Hern{\'a}ndez-Toledo}
  H.~M.,  {V{\'a}zquez-Mata} J.~A.,   {S{\'a}nchez} S.~F.,  2020, arXiv
  e-prints, \href {https://ui.adsabs.harvard.edu/abs/2020arXiv200105506L} {p.
  arXiv:2001.05506}

\bibitem[\protect\citeauthoryear{{Lacey} et~al.,}{{Lacey}
  et~al.}{2016}]{Durham2016}
{Lacey} C.~G.,  et~al., 2016, \mn@doi [\mnras] {10.1093/mnras/stw1888}, \href
  {https://ui.adsabs.harvard.edu/abs/2016MNRAS.462.3854L} {462, 3854}

\bibitem[\protect\citeauthoryear{{Law} et~al.,}{{Law} et~al.}{2015}]{Law2015}
{Law} D.~R.,  et~al., 2015, \mn@doi [\aj] {10.1088/0004-6256/150/1/19}, \href
  {https://ui.adsabs.harvard.edu/abs/2015AJ....150...19L} {150, 19}

\bibitem[\protect\citeauthoryear{{Law} et~al.,}{{Law} et~al.}{2016}]{Law2016}
{Law} D.~R.,  et~al., 2016, \mn@doi [\aj] {10.3847/0004-6256/152/4/83}, \href
  {https://ui.adsabs.harvard.edu/abs/2016AJ....152...83L} {152, 83}

\bibitem[\protect\citeauthoryear{{Li} et~al.,}{{Li} et~al.}{2017}]{Li17}
{Li} H.,  et~al., 2017, \mn@doi [\apj] {10.3847/1538-4357/aa662a}, \href
  {http://adsabs.harvard.edu/abs/2017ApJ...838...77L} {838, 77}

\bibitem[\protect\citeauthoryear{{Li} et~al.,}{{Li} et~al.}{2018}]{Li2018}
{Li} H.,  et~al., 2018, \mn@doi [\mnras] {10.1093/mnras/sty334}, \href
  {https://ui.adsabs.harvard.edu/abs/2018MNRAS.476.1765L} {476, 1765}

\bibitem[\protect\citeauthoryear{{Liang}, {Li}, {Li}, {Zhou}, {Yan}, {Mo}  \&
  {Zhang}}{{Liang} et~al.}{2021}]{Liang2021}
{Liang} F.-H.,  {Li} C.,  {Li} N.,  {Zhou} S.,  {Yan} R.,  {Mo} H.,   {Zhang}
  W.,  2021, \mn@doi [\apj] {10.3847/1538-4357/ac2bff}, \href
  {https://ui.adsabs.harvard.edu/abs/2021ApJ...923..120L} {923, 120}

\bibitem[\protect\citeauthoryear{{Marsden}, {Shankar}, {Bernardi}, {Sheth},
  {Fu}  \& {Lapi}}{{Marsden} et~al.}{2021}]{Marsden2022}
{Marsden} C.,  {Shankar} F.,  {Bernardi} M.,  {Sheth} R.,  {Fu} H.,   {Lapi}
  A.,  2021, arXiv e-prints, \href
  {https://ui.adsabs.harvard.edu/abs/2021arXiv211209720M} {p. arXiv:2112.09720}

\bibitem[\protect\citeauthoryear{{Mart{\'\i}n-Navarro}, {La Barbera},
  {Vazdekis}, {Falc{\'o}n-Barroso}  \& {Ferreras}}{{Mart{\'\i}n-Navarro}
  et~al.}{2015}]{MartinNavarro2015}
{Mart{\'\i}n-Navarro} I.,  {La Barbera} F.,  {Vazdekis} A.,
  {Falc{\'o}n-Barroso} J.,   {Ferreras} I.,  2015, \mn@doi [\mnras]
  {10.1093/mnras/stu2480}, \href
  {https://ui.adsabs.harvard.edu/abs/2015MNRAS.447.1033M} {447, 1033}

\bibitem[\protect\citeauthoryear{{Mart{\'{\i}}n-Navarro}, {Vazdekis},
  {Falc{\'o}n-Barroso}, {La Barbera}, {Y{\i}ld{\i}r{\i}m}  \& {van de
  Ven}}{{Mart{\'{\i}}n-Navarro} et~al.}{2018}]{LaBarbera2018}
{Mart{\'{\i}}n-Navarro} I.,  {Vazdekis} A.,  {Falc{\'o}n-Barroso} J.,  {La
  Barbera} F.,  {Y{\i}ld{\i}r{\i}m} A.,   {van de Ven} G.,  2018, \mn@doi
  [\mnras] {10.1093/mnras/stx3346}, \href
  {http://adsabs.harvard.edu/abs/2018MNRAS.475.3700M} {475, 3700}

\bibitem[\protect\citeauthoryear{{Mehlert}, {Thomas}, {Saglia}, {Bender}  \&
  {Wegner}}{{Mehlert} et~al.}{2003}]{Mehlert2003}
{Mehlert} D.,  {Thomas} D.,  {Saglia} R.~P.,  {Bender} R.,   {Wegner} G.,
  2003, \mn@doi [\aap] {10.1051/0004-6361:20030886}, \href
  {https://ui.adsabs.harvard.edu/abs/2003A&A...407..423M} {407, 423}

\bibitem[\protect\citeauthoryear{{Mendel}, {Simard}, {Palmer}, {Ellison}  \&
  {Patton}}{{Mendel} et~al.}{2014}]{Mendel2014}
{Mendel} J.~T.,  {Simard} L.,  {Palmer} M.,  {Ellison} S.~L.,   {Patton} D.~R.,
   2014, \mn@doi [\apjs] {10.1088/0067-0049/210/1/3}, \href
  {http://adsabs.harvard.edu/abs/2014ApJS..210....3M} {210, 3}

\bibitem[\protect\citeauthoryear{{Mendel} et~al.,}{{Mendel}
  et~al.}{2020}]{Mendel2020}
{Mendel} J.~T.,  et~al., 2020, \mn@doi [\apj] {10.3847/1538-4357/ab9ffc}, \href
  {https://ui.adsabs.harvard.edu/abs/2020ApJ...899...87M} {899, 87}

\bibitem[\protect\citeauthoryear{{Mowla} et~al.,}{{Mowla}
  et~al.}{2019}]{Mowla2019}
{Mowla} L.~A.,  et~al., 2019, \mn@doi [\apj] {10.3847/1538-4357/ab290a}, \href
  {https://ui.adsabs.harvard.edu/abs/2019ApJ...880...57M} {880, 57}

\bibitem[\protect\citeauthoryear{{Nelson} et~al.,}{{Nelson}
  et~al.}{2016}]{Nelson2016}
{Nelson} E.~J.,  et~al., 2016, \mn@doi [\apj] {10.3847/0004-637X/828/1/27},
  \href {https://ui.adsabs.harvard.edu/abs/2016ApJ...828...27N} {828, 27}

\bibitem[\protect\citeauthoryear{{Parikh} et~al.,}{{Parikh}
  et~al.}{2018}]{Parikh2018}
{Parikh} T.,  et~al., 2018, \mn@doi [\mnras] {10.1093/mnras/sty785}, \href
  {http://adsabs.harvard.edu/abs/2018MNRAS.477.3954P} {477, 3954}

\bibitem[\protect\citeauthoryear{{Pietrinferni}, {Cassisi}, {Salaris}  \&
  {Castelli}}{{Pietrinferni} et~al.}{2006}]{Pietrinferni2006}
{Pietrinferni} A.,  {Cassisi} S.,  {Salaris} M.,   {Castelli} F.,  2006,
  \mn@doi [\apj] {10.1086/501344}, \href
  {http://adsabs.harvard.edu/abs/2006ApJ...642..797P} {642, 797}

\bibitem[\protect\citeauthoryear{{Pietrinferni}, {Cassisi}, {Salaris}  \&
  {Hidalgo}}{{Pietrinferni} et~al.}{2013}]{Pietrinferni2013}
{Pietrinferni} A.,  {Cassisi} S.,  {Salaris} M.,   {Hidalgo} S.,  2013, \mn@doi
  [\aap] {10.1051/0004-6361/201321950}, \href
  {http://adsabs.harvard.edu/abs/2013A%26A...558A..46P} {558, A46}

\bibitem[\protect\citeauthoryear{{Santucci} et~al.,}{{Santucci}
  et~al.}{2020}]{Santucci2020}
{Santucci} G.,  et~al., 2020, \mn@doi [\apj] {10.3847/1538-4357/ab92a9}, \href
  {https://ui.adsabs.harvard.edu/abs/2020ApJ...896...75S} {896, 75}

\bibitem[\protect\citeauthoryear{{S{\'e}rsic}}{{S{\'e}rsic}}{1963}]{Sersic1963}
{S{\'e}rsic} J.~L.,  1963, Boletin de la Asociacion Argentina de Astronomia La
  Plata Argentina, \href {http://adsabs.harvard.edu/abs/1963BAAA....6...41S}
  {6, 41}

\bibitem[\protect\citeauthoryear{{Shankar}, {Marulli}, {Bernardi}, {Mei},
  {Meert}  \& {Vikram}}{{Shankar} et~al.}{2013}]{Shankar2013}
{Shankar} F.,  {Marulli} F.,  {Bernardi} M.,  {Mei} S.,  {Meert} A.,   {Vikram}
  V.,  2013, \mn@doi [\mnras] {10.1093/mnras/sts001}, \href
  {https://ui.adsabs.harvard.edu/abs/2013MNRAS.428..109S} {428, 109}

\bibitem[\protect\citeauthoryear{{Shankar} et~al.,}{{Shankar}
  et~al.}{2015}]{Shankar2015}
{Shankar} F.,  et~al., 2015, \mn@doi [\apj] {10.1088/0004-637X/802/2/73}, \href
  {https://ui.adsabs.harvard.edu/abs/2015ApJ...802...73S} {802, 73}

\bibitem[\protect\citeauthoryear{{Shankar} et~al.,}{{Shankar}
  et~al.}{2018}]{Shankar2018}
{Shankar} F.,  et~al., 2018, \mn@doi [\mnras] {10.1093/mnras/stx3086}, \href
  {https://ui.adsabs.harvard.edu/abs/2018MNRAS.475.2878S} {475, 2878}

\bibitem[\protect\citeauthoryear{{Sharda} \& {Krumholz}}{{Sharda} \&
  {Krumholz}}{2022}]{Sharda2022}
{Sharda} P.,  {Krumholz} M.~R.,  2022, \mn@doi [\mnras]
  {10.1093/mnras/stab2921}, \href
  {https://ui.adsabs.harvard.edu/abs/2022MNRAS.509.1959S} {509, 1959}

\bibitem[\protect\citeauthoryear{{Shen}, {Mo}, {White}, {Blanton}, {Kauffmann},
  {Voges}, {Brinkmann}  \& {Csabai}}{{Shen} et~al.}{2003}]{Shen2003}
{Shen} S.,  {Mo} H.~J.,  {White} S. D.~M.,  {Blanton} M.~R.,  {Kauffmann} G.,
  {Voges} W.,  {Brinkmann} J.,   {Csabai} I.,  2003, \mn@doi [\mnras]
  {10.1046/j.1365-8711.2003.06740.x}, \href
  {https://ui.adsabs.harvard.edu/abs/2003MNRAS.343..978S} {343, 978}

\bibitem[\protect\citeauthoryear{{Smee} et~al.,}{{Smee}
  et~al.}{2013}]{Smee2013}
{Smee} S.~A.,  et~al., 2013, \mn@doi [\aj] {10.1088/0004-6256/146/2/32}, \href
  {https://ui.adsabs.harvard.edu/abs/2013AJ....146...32S} {146, 32}

\bibitem[\protect\citeauthoryear{{Smith}}{{Smith}}{2020}]{Smith2020}
{Smith} R.~J.,  2020, \mn@doi [\araa] {10.1146/annurev-astro-032620-020217},
  \href {https://ui.adsabs.harvard.edu/abs/2020ARA&A..58..577S} {58, 577}

\bibitem[\protect\citeauthoryear{{Spolaor}, {Proctor}, {Forbes}  \&
  {Couch}}{{Spolaor} et~al.}{2009}]{Spolaor2009}
{Spolaor} M.,  {Proctor} R.~N.,  {Forbes} D.~A.,   {Couch} W.~J.,  2009,
  \mn@doi [\apjl] {10.1088/0004-637X/691/2/L138}, \href
  {https://ui.adsabs.harvard.edu/abs/2009ApJ...691L.138S} {691, L138}

\bibitem[\protect\citeauthoryear{{Suess}, {Kriek}, {Price}  \& {Barro}}{{Suess}
  et~al.}{2019}]{Suess2019}
{Suess} K.~A.,  {Kriek} M.,  {Price} S.~H.,   {Barro} G.,  2019, \mn@doi [\apj]
  {10.3847/1538-4357/ab1bda}, \href
  {https://ui.adsabs.harvard.edu/abs/2019ApJ...877..103S} {877, 103}

\bibitem[\protect\citeauthoryear{{Szomoru}, {Franx}, {van Dokkum}, {Trenti},
  {Illingworth}, {Labb{\'e}}  \& {Oesch}}{{Szomoru} et~al.}{2013}]{Szomoru2013}
{Szomoru} D.,  {Franx} M.,  {van Dokkum} P.~G.,  {Trenti} M.,  {Illingworth}
  G.~D.,  {Labb{\'e}} I.,   {Oesch} P.,  2013, \mn@doi [\apj]
  {10.1088/0004-637X/763/2/73}, \href
  {https://ui.adsabs.harvard.edu/abs/2013ApJ...763...73S} {763, 73}

\bibitem[\protect\citeauthoryear{{Tinsley}}{{Tinsley}}{1972}]{Tinsley1972}
{Tinsley} B.~M.,  1972, \mn@doi [\apj] {10.1086/151793}, \href
  {https://ui.adsabs.harvard.edu/abs/1972ApJ...178..319T} {178, 319}

\bibitem[\protect\citeauthoryear{{Tortora}, {Napolitano}, {Romanowsky},
  {Jetzer}, {Cardone}  \& {Capaccioli}}{{Tortora} et~al.}{2011}]{Tortora2011}
{Tortora} C.,  {Napolitano} N.~R.,  {Romanowsky} A.~J.,  {Jetzer} P.,
  {Cardone} V.~F.,   {Capaccioli} M.,  2011, \mn@doi [\mnras]
  {10.1111/j.1365-2966.2011.19438.x}, \href
  {http://adsabs.harvard.edu/abs/2011MNRAS.418.1557T} {418, 1557}

\bibitem[\protect\citeauthoryear{{Vaughan}, {Davies}, {Zieleniewski}  \&
  {Houghton}}{{Vaughan} et~al.}{2018}]{Vaughan18}
{Vaughan} S.~P.,  {Davies} R.~L.,  {Zieleniewski} S.,   {Houghton} R.~C.~W.,
  2018, \mn@doi [\mnras] {10.1093/mnras/sty1434}, \href
  {http://adsabs.harvard.edu/abs/2018MNRAS.479.2443V} {479, 2443}

\bibitem[\protect\citeauthoryear{{Vazdekis}, {Casuso}, {Peletier}  \&
  {Beckman}}{{Vazdekis} et~al.}{1996}]{Vazdekis1996}
{Vazdekis} A.,  {Casuso} E.,  {Peletier} R.~F.,   {Beckman} J.~E.,  1996,
  \mn@doi [\apjs] {10.1086/192340}, \href
  {http://adsabs.harvard.edu/abs/1996ApJS..106..307V} {106, 307}

\bibitem[\protect\citeauthoryear{{Vazdekis} et~al.,}{{Vazdekis}
  et~al.}{2015}]{Vazdekis2015}
{Vazdekis} A.,  et~al., 2015, \mn@doi [\mnras] {10.1093/mnras/stv151}, \href
  {https://ui.adsabs.harvard.edu/abs/2015MNRAS.449.1177V} {449, 1177}

\bibitem[\protect\citeauthoryear{{Wake} et~al.,}{{Wake}
  et~al.}{2017}]{Wake2017}
{Wake} D.~A.,  et~al., 2017, \mn@doi [\aj] {10.3847/1538-3881/aa7ecc}, \href
  {http://adsabs.harvard.edu/abs/2017AJ....154...86W} {154, 86}

\bibitem[\protect\citeauthoryear{{Westfall} et~al.,}{{Westfall}
  et~al.}{2019}]{Westfall2019}
{Westfall} K.~B.,  et~al., 2019, \mn@doi [\aj] {10.3847/1538-3881/ab44a2},
  \href {https://ui.adsabs.harvard.edu/abs/2019AJ....158..231W} {158, 231}

\bibitem[\protect\citeauthoryear{{Yan} et~al.,}{{Yan} et~al.}{2016a}]{Yan2016a}
{Yan} R.,  et~al., 2016a, \mn@doi [\aj] {10.3847/0004-6256/151/1/8}, \href
  {https://ui.adsabs.harvard.edu/abs/2016AJ....151....8Y} {151, 8}

\bibitem[\protect\citeauthoryear{{Yan} et~al.,}{{Yan} et~al.}{2016b}]{Yan2016b}
{Yan} R.,  et~al., 2016b, \mn@doi [\aj] {10.3847/0004-6256/152/6/197}, \href
  {https://ui.adsabs.harvard.edu/abs/2016AJ....152..197Y} {152, 197}

\bibitem[\protect\citeauthoryear{{Yan}, {Je{\v{r}}{\'a}bkov{\'a}}  \&
  {Kroupa}}{{Yan} et~al.}{2021}]{Yan2021}
{Yan} Z.,  {Je{\v{r}}{\'a}bkov{\'a}} T.,   {Kroupa} P.,  2021, \mn@doi [\aap]
  {10.1051/0004-6361/202140683}, \href
  {https://ui.adsabs.harvard.edu/abs/2021A&A...655A..19Y} {655, A19}

\bibitem[\protect\citeauthoryear{{Zanisi} et~al.,}{{Zanisi}
  et~al.}{2021}]{Zanisi2021}
{Zanisi} L.,  et~al., 2021, \mn@doi [\mnras] {10.1093/mnras/stab1472}, \href
  {https://ui.adsabs.harvard.edu/abs/2021MNRAS.505.4555Z} {505, 4555}

\bibitem[\protect\citeauthoryear{{de Graaff} et~al.,}{{de Graaff}
  et~al.}{2021}]{deGraaff2021}
{de Graaff} A.,  et~al., 2021, \mn@doi [\apj] {10.3847/1538-4357/abf1e7}, \href
  {https://ui.adsabs.harvard.edu/abs/2021ApJ...913..103D} {913, 103}

\bibitem[\protect\citeauthoryear{{de Vaucouleurs}}{{de
  Vaucouleurs}}{1959}]{Ttype1959}
{de Vaucouleurs} G.,  1959, \mn@doi [Handbuch der Physik]
  {10.1007/978-3-642-45932-0\_7}, \href
  {https://ui.adsabs.harvard.edu/abs/1959HDP....53..275D} {53, 275}

\bibitem[\protect\citeauthoryear{{van Dokkum} \& {Franx}}{{van Dokkum} \&
  {Franx}}{1996}]{vanDokkumFranx1996}
{van Dokkum} P.~G.,  {Franx} M.,  1996, \mn@doi [\mnras]
  {10.1093/mnras/281.3.985}, \href
  {https://ui.adsabs.harvard.edu/abs/1996MNRAS.281..985V} {281, 985}

\bibitem[\protect\citeauthoryear{{van Dokkum} et~al.,}{{van Dokkum}
  et~al.}{2008}]{vanDokkum2008}
{van Dokkum} P.~G.,  et~al., 2008, \mn@doi [\apjl] {10.1086/587874}, \href
  {https://ui.adsabs.harvard.edu/abs/2008ApJ...677L...5V} {677, L5}

\bibitem[\protect\citeauthoryear{{van Dokkum} et~al.,}{{van Dokkum}
  et~al.}{2010}]{vD2010}
{van Dokkum} P.~G.,  et~al., 2010, \mn@doi [\apj]
  {10.1088/0004-637X/709/2/1018}, \href
  {https://ui.adsabs.harvard.edu/abs/2010ApJ...709.1018V} {709, 1018}

\bibitem[\protect\citeauthoryear{{van Dokkum}, {Conroy}, {Villaume}, {Brodie}
  \& {Romanowsky}}{{van Dokkum} et~al.}{2017}]{vD2017}
{van Dokkum} P.,  {Conroy} C.,  {Villaume} A.,  {Brodie} J.,   {Romanowsky}
  A.~J.,  2017, \mn@doi [\apj] {10.3847/1538-4357/aa7135}, \href
  {http://adsabs.harvard.edu/abs/2017ApJ...841...68V} {841, 68}

\bibitem[\protect\citeauthoryear{{van der Wel}, {Bell}, {van den Bosch},
  {Gallazzi}  \& {Rix}}{{van der Wel} et~al.}{2009}]{vanderWel2009}
{van der Wel} A.,  {Bell} E.~F.,  {van den Bosch} F.~C.,  {Gallazzi} A.,
  {Rix} H.-W.,  2009, \mn@doi [\apj] {10.1088/0004-637X/698/2/1232}, \href
  {https://ui.adsabs.harvard.edu/abs/2009ApJ...698.1232V} {698, 1232}

\bibitem[\protect\citeauthoryear{{van der Wel} et~al.,}{{van der Wel}
  et~al.}{2014}]{vanderWel2014}
{van der Wel} A.,  et~al., 2014, \mn@doi [\apj] {10.1088/0004-637X/788/1/28},
  \href {https://ui.adsabs.harvard.edu/abs/2014ApJ...788...28V} {788, 28}

\makeatother
\end{thebibliography}





\bsp	
\label{lastpage}
\end{document}